\documentclass[a4paper,fleqn]{cas-dc}
\usepackage[numbers]{natbib}
\usepackage{times}
\usepackage{epsfig}
\usepackage{graphicx}
\usepackage{amsmath}
\usepackage{amssymb}
\usepackage{multirow}
\usepackage{subfigure}
\usepackage{comment}
\newcommand{\minisection}[1]{\vspace{0.04in} \noindent {\bf #1}}

\def\tsc#1{\csdef{#1}{\textsc{\lowercase{#1}}\xspace}}
\tsc{WGM}
\tsc{QE}

\begin{document}
\let\WriteBookmarks\relax
\def\floatpagepagefraction{1}
\def\textpagefraction{.001}

\shorttitle{A Novel Framework for Image-to-image Translation and Image Compression}    

\shortauthors{Fei Yang, Yaxing Wang, Luis Herranz, Yongmei Cheng, Mikhail Mozerov}  

\title [mode = title]{A Novel Framework for Image-to-image Translation and Image Compression}  

\author[1,2]{Fei Yang}[]
\cormark[1]
\cortext[1]{Corresponding author}
\ead{fyang@cvc.uab.cat}
\author[2]{Yaxing Wang}[]
\author[2]{Luis Herranz}[]
\author[1]{Yongmei Cheng}[]
\author[2]{Mikhail G. Mozerov}[]


\affiliation[1]{organization={School of Automation},
            addressline={Northwestern Polytechnical University}, 
            city={Xi'an},
            postcode={710129}, 
            country={China}}
            
\affiliation[2]{organization={Computer Vision Center},
            addressline={Universitat Autònoma de Barcelona}, 
            city={Barcelona},
            postcode={08192}, 
            country={Spain}}

\begin{abstract}
Data-driven paradigms using machine learning are becoming ubiquitous in image processing and communications. In particular, image-to-image (I2I) translation is a generic and widely used approach to image processing problems, such as image synthesis, style transfer, and image restoration. At the same time, neural image compression has emerged as a data-driven alternative to traditional coding approaches in visual communications. In this paper, we study the combination of these two paradigms into a joint I2I compression and translation framework, focusing on multi-domain image synthesis. We first propose distributed I2I translation by integrating quantization and entropy coding into an I2I translation framework (i.e. I2Icodec). In practice, the image compression functionality (i.e. autoencoding) is also desirable, requiring to deploy alongside I2Icodec a regular image codec. Thus, we further propose a unified framework that allows both translation and autoencoding capabilities in a single codec. Adaptive residual blocks conditioned on the translation/compression mode provide flexible adaptation to the desired functionality. The experiments show promising results in both I2I translation and image compression
using a single model.
\end{abstract}

\begin{keywords}
 \sep Image-to-image translation \sep Autoencoder \sep Image compression \sep Communication
\end{keywords}

\maketitle

\section{Introduction}
\label{sec:intro}
Modern computer vision and image processing heavily rely on deep neural networks and machine learning. One prominent example is image-to-image (I2I) translation, which addresses the problem of learning to transform images from a source domain to a target domain. This general approach has numerous applications in image restoration and enhancement (e.g. colorization, superresolution, deblurring), but also more complex data-driven transformations where the transformation is learned from data (e.g. style transfer, face attribute modification, scene synthesis, zebra-to-horse translation). Recently, deep neural networks have been also applied to image coding, resulting  in the alternative coding paradigm of neural image compression (NIC). These approaches can compete and often surpasse the rate-distortion performance of traditional transform coding approaches (e.g. BPG~\cite{bpg}). Image and video coding technology has also significant implications in visual communications and the storage and distribution of video content.

Building upon those two aforemention research areas, in this paper we study the problem of \textit{distributed I2I translation}, where encoding is performed at the sender side, decoding at the receiver side and the coded representation is either transmitted through a digital communications channel or stored. Thus, in addition to addressing the translation problem, we also aim at obtaining compact binary representations (i.e. bitstreams).

\begin{figure*}[!t]
	\centering
	\includegraphics[width=\textwidth]{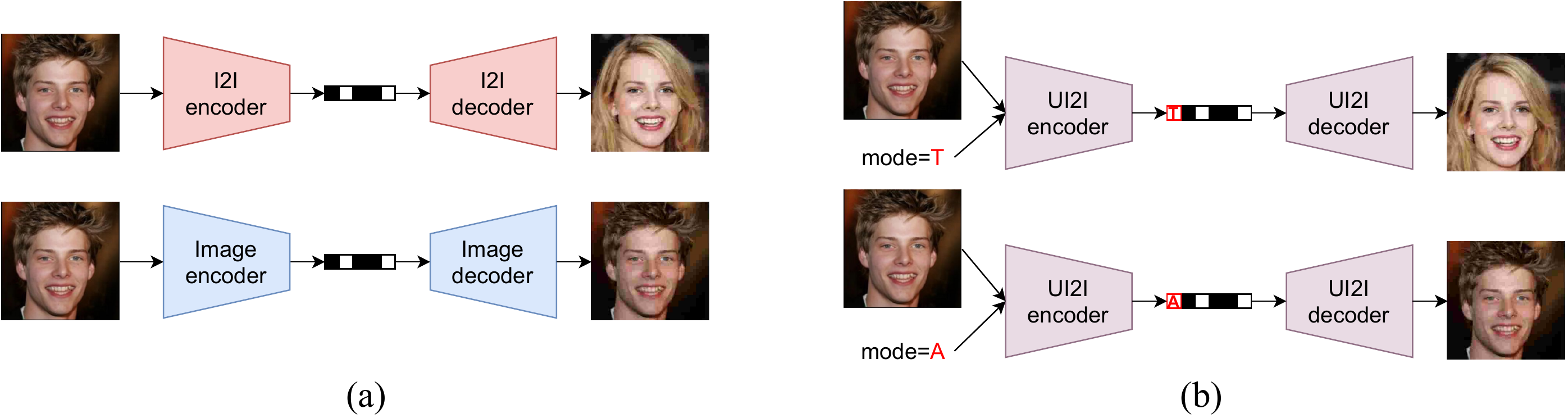}
	\caption{Proposed approaches: (a) distributed I2I translation (\textit{I2Icodec}), alongside a regular image compression codec, and (b) unified translation and autoencoding framework using a single codec (\textit{UI2Icodec}).}
	\label{fig:motivation}
\end{figure*}

A naive approach to this problem would be translating the image at the sender side and then compressing the result, or translating the reconstructed image at the receiver side. These approaches have several limitations. First, they require encoding and decoding images twice, once with the translator and once with the image codec, resulting in lower computational efficiency. Similarly, it requires storing two separate codecs (i.e. for translation and autoencoding). Finally, each encoding and decoding pass is a lossy transformation, therefore it is likely that through these four transformations more information is lost, resulting in lower quality in the translation with artifacts, and/or larger bitstreams. In order to address these limitations, we propose the I2Icodec framework (see Figure~\ref{fig:motivation}a), which addresses distributed I2I translation with a single encoder and decoder, thus avoiding computational overheads and potential loss of information.

While I2Icodec only requires a single encoder and decoder pair, it cannot perform regular autoencoding (i.e. conventional image coding), which is a desirable functionality in practice. A naive solution is to deploy a regular image codec alongside, but that increases the memory requirements significantly. Thus, to avoid deploying two separate models, we also propose UI2Icodec, a unified framework that can perform both distributed translation and autoencoding in a single model (see Figure~\ref{fig:motivation}b).

Note that distributed I2I translation is not limited to \textit{spatially} dislocated encoder and decoder, but it could also be applied to store local files that will be decoded in the future (i.e. \textit{temporally} distributed I2I translation). In this way, a single compact model prevents unnecessary use of computation and memory resources.

In summary, our contributions are: 1) we study the problem of distributed I2I translation, which involves I2I translation under rate constraints; 2) a novel framework for distributed I2I translation (I2Icodec); and 3) a unified framework for distributed I2I translation and autoencoding (UI2Icodec).

\section{Related work}
\subsection{Image-to-image translation}
Image-to-image translation has been studied widely in recent years. Paired I2I translation~\cite{isola2016image,zhu2017toward,gonzalez2018image,wang2018video,wang2018fewshotvid2vid} assumes the availability of input-output image pairs. Unpaired I2I translation~\cite{liu2017unsupervised,kim2017learning,yi2017dualgan,zhu2017unpaired,gan2017triangle,mejjati2018unsupervised,park2020contrastive} is a more challenging setting where translations are learned from (unpaired) images from the input and output domain. 

Often, a given input image can have multiple plausible translations (e.g. colorization of grayscale images). Multimodal I2I translation (or diverse I2I translation)~\cite{huang2018multimodal,Lee2018drit,almahairi2018augmented,royer2018xgan,yu2019multi} addresses this problem by disentangling content and style. Style is sampled randomly, ensuring the model generates diverse translations.
Early approaches assume only two domains. More recently, multi-domain I2I translation approaches~\cite{ yu2019multi,choi2020stargan,lee2020drit++} can translate between a range of domains using a single model. In particular, we build upon on StarGAN v2~\cite{choi2020stargan} which provides state-of-the-art translation, including multimodal and multi-domain translation, with content-style disentanglement.

\subsection{Neural image compression}
Motivated by the success of deep learning, neural image compression~\cite{toderici2015variable, balle2016end, theis2017lossy} has emerged as a new paradigm where codecs are implemented as deep neural networks. Their parameters are directly optimized to minimize a particular combination of rate and distortion over a training dataset, a clear advantage over traditionally engineered transform coding. One key obstacle is quantization and entropy coding, which are non-differentiable operations. In practice, during training, quantization is replaced by a differentiable proxy~\cite{balle2016end, theis2017lossy}, or soft vector quantization~\cite{agustsson2017soft}, and entropy coding is bypassed, with rate estimated as the entropy of the quantized latent representation. The architecture is based on convolutional autoencoder~\cite{balle2016end, theis2017lossy}, sometimes with recurrent neural networks~\cite{toderici2015variable, toderici2017full}. A key component in minimizing the rate is the learnable entropy model. Recent models include hyperpriors~\cite{balle2018variational}, autoregressive models~\cite{minnen2018joint, lee2018context, mentzer2018conditional, li2020efficient, minnen2020channel} and generative models~\cite{agustsson2019generative, blau2019rethinking}. More recently, variable-rate approaches enable encoding at multiple rate-distortion tradeoffs within the same model~\cite{theis2017lossy,choi2019variable,yang2020variable}. Aiming at decoding realistic reconstructions even with low rates, some works~\cite{patel2019deep, patel2020hierarchical, patel2021saliency, mentzer2020high} explore the use of perceptual and adversarial losses during training. Motivated by this, we also use adversarial loss for both translation and autoencoding.

\section{Distributed image-to-image translation}

We first consider the problem of distributed I2I translation to a target domain. In particular, an image $\mathbf{x}\in \mathcal{X}$ from a source domain label $y_{src}\in \mathcal{Y}_{src}$ is transformed in a compact latent representation $\mathbf{z}$ and encoded into a bitstream $\mathbf{b}$ at the sender side. Following the common practice of disentangling content and style, the expected style of the translated image should also be sent to the receiver side while $\mathbf{z}$ corresponds to the content. Then, the receiver side can decode $\mathbf{b}$, reconstructing the translated image in a target domain, indicated by the label $y_{tar}\in \mathcal{Y}_{tar}$. The style of the translated image is determined by a style vector $\mathbf{s}$, either sampled randomly or obtained from a reference image. This disentanglement enables the reconstruction of diverse translations for a given image $\mathbf{x}$. The style vector $\mathbf{s}$ can be provided by either the sender or the receiver. In the former, the style vector is quantized and included in the bitstream $\mathbf{b}$ and transmitted or stored accordingly (being very compact, the overhead is negligible).

The objective in distributed I2I translation is to obtain successful translations with compact bitstreams.

\subsection{I2Icodec framework}
Our framework is based on the I2I translation framework of \cite{choi2020stargan}, but with the encoder and decoder located separately in the sender and receiver sides, respectively (see Figure 2). In order to achieve this, the framework is augmented with compression capabilities, i.e. quantization and entropy coding. It is composed of \textit{content encoder} $E^{c}$,  \textit{style encoder} $E^{s}$,  \textit{mapping network} $M$, \textit{decoder} $G$  and  \textit{discriminator} $D^T$.

\minisection{Content encoder} The content encoder $E^c$ extracts a latent representation $\mathbf{z}=E^{c}\left(\mathbf{x}\right)$ of the content of the image $\mathbf{x}$. To transmit this representation via a binary channel, the representation $\mathbf{z}$ is quantized (in this paper we use scalar quantization) as $\mathbf{q}=Q\left(\mathbf{z}\right)$ to obtain a discrete-valued representation $\mathbf{q}\in \mathbb{Z}^D$. Then, $\mathbf{q}$ is binarized using entropy coding (arithmetic coding in our case) to reduce statistical redundancy. Quantization is lossy, but entropy coding is not. During training, we replace quantization by uniform noise, and bypass entropy coding, approximating the rate by the entropy of $\mathbf{z}$.

\minisection{Mapping network and style encoder.} The style $\mathbf{s}$ is used for guiding I2I translation towards a specific style in the target domain. This style code can either be sampled randomly (providing diversity), or obtained from a reference style image. In the former, the mapping network $M$ obtains the domain-specific style representation $\mathbf{s}$ from a domain-independent random style $\mathbf{w}$ as $\mathbf{s}=M\left(\mathbf{w},y_{tar}\right)$. Alternatively, the domain-specific style representation can be obtained from a reference image $\mathbf{\bar{x}}$ with the style encoder $E^{s}$ as $\mathbf{s}=E^{s}\left(\mathbf{\bar{x}},y_{tar}\right)$. 

\minisection{Decoder.} The decoder $G$ receives the bitstream and performs entropy decoding and maps back to the real-valued representation $\hat{\mathbf{z}}$. It then generates the reconstructed image from $\mathbf{z}$ and the style $\mathbf{s}$ as $\hat{\mathbf{x}}=G\left(\hat{\mathbf{z}},\mathbf{s}\right)$. 

\minisection{Discriminator.} Following \cite{choi2020stargan}, we use a multi-task discriminator where $D^{T}\left(\hat{\mathbf{x}},y_{tar}\right)$ returns the probability that $\hat{\mathbf{x}}$ is classified in target domain $y_{tar}$. 

\minisection{Entropy model.} We use a learnable hyperprior~\cite{balle2018variational} to model the latent distribution $P(\mathbf{z})$.

\begin{figure*}[!t]
  \centering
  \includegraphics[width=\textwidth]{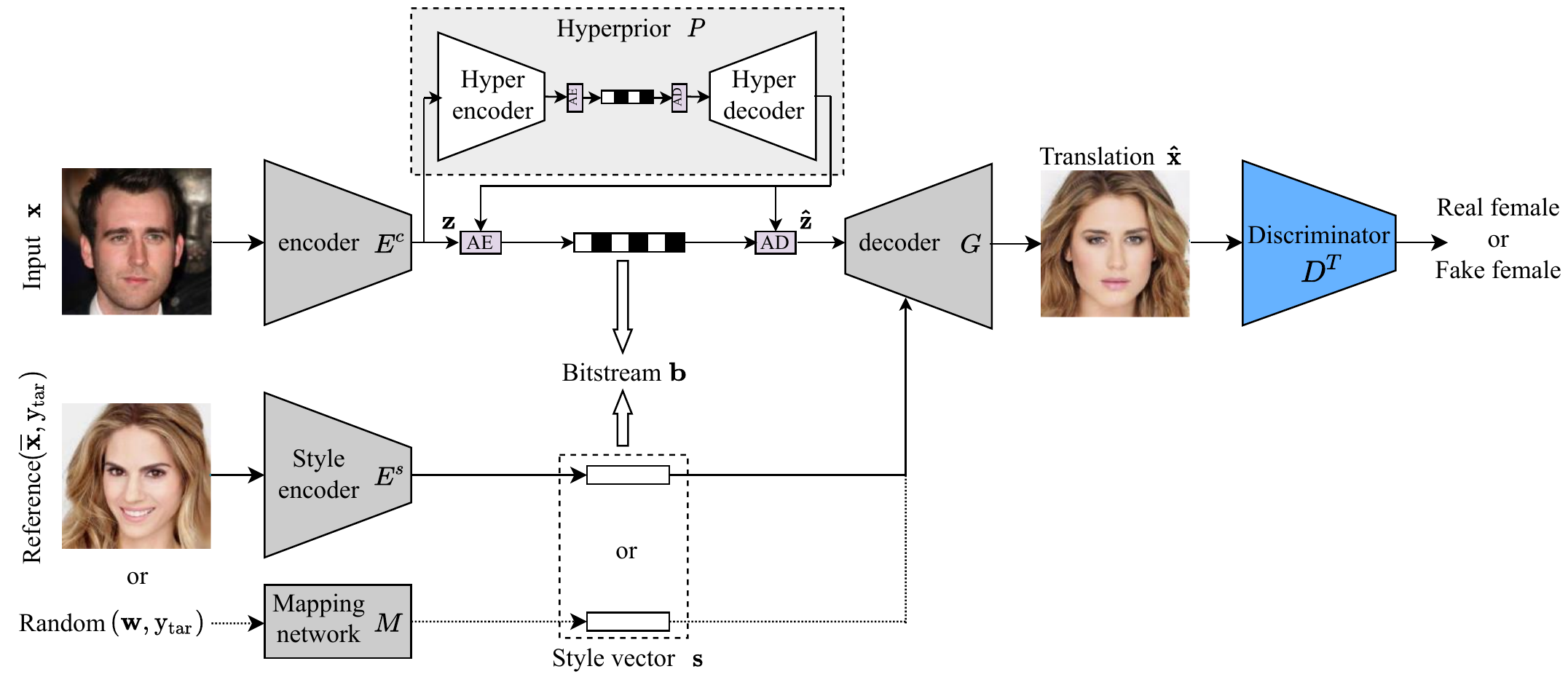}
  \vspace{-1.2em}
  \caption{Architecture of the framework for distributed I2I translation (\textit{I2Icodec}). The style vector $\mathbf{s}$ needs to be included in the bitstream when it is provided on the sender side and not when on the receiver side.}
  \label{model_I2I}
  \vspace{-1.2em}
\end{figure*}

\begin{figure*}[!t]
  \centering
  \includegraphics[width=\textwidth]{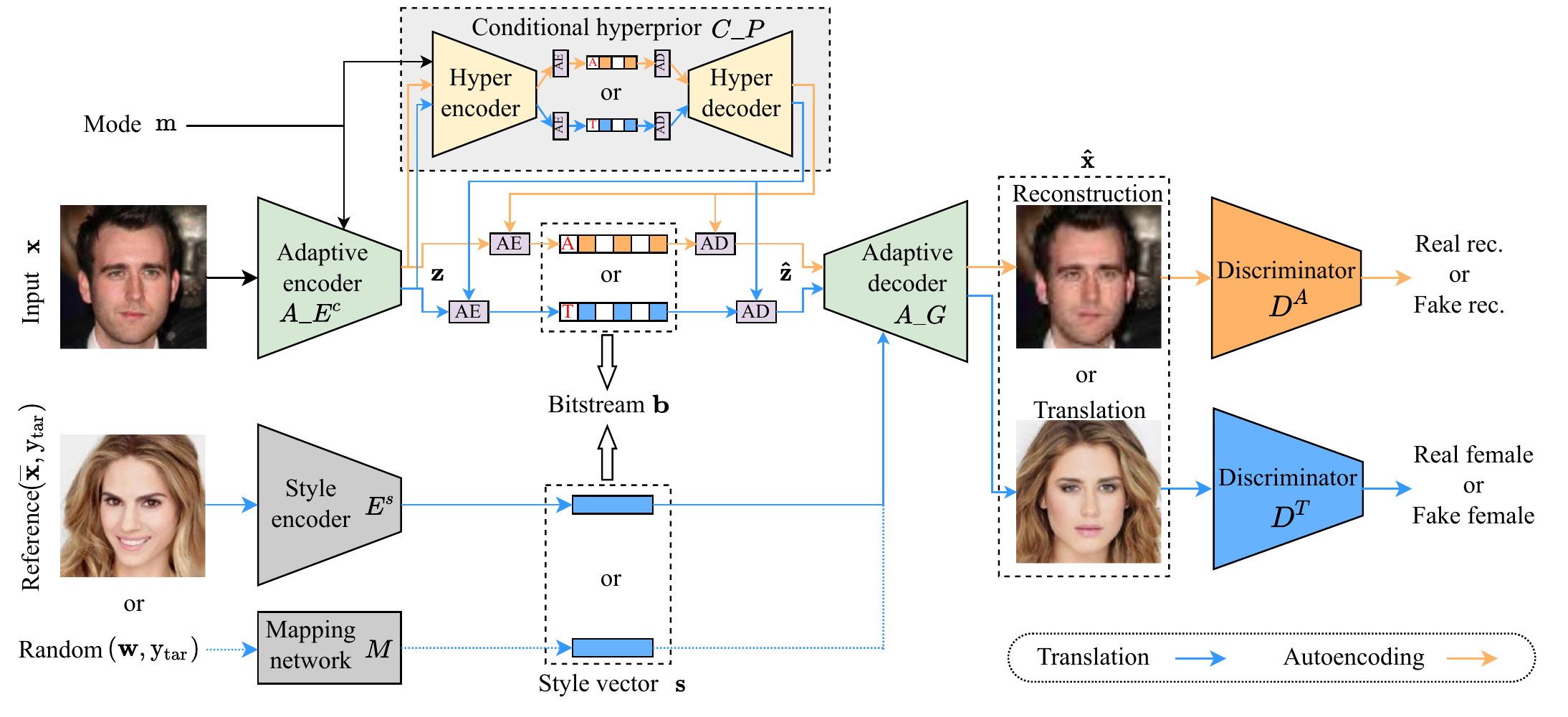}
  \vspace{-1.2em}
  \caption{Architecture of the unified framework for regular image compression and I2I translation (\textit{UI2Icodec}). The mode $m$ signals whether the model runs as autoencoder or as I2I translator (store as 1 bit in the bitstream).}
  \label{model}
  \vspace{-1.2em}
\end{figure*}

\subsection{Loss}
Our objective during training is to optimize translation while minimizing the rate. Regarding translation, we follow the losses used in~\cite{choi2020stargan}. Given an image $\mathbf{x} \in \cal{X}$ and its original domain label $y_{src} \in \mathcal{Y}_{src}$, we can obtain its latent representation $\mathbf{z}=E^c(\mathbf{x})$. The I2I translation loss $\mathcal{L}_{T}$ consists of the following terms.

\minisection{Adversarial loss.} In principle, the adversarial loss forces the translated images to be indistinguishable from real photos. We first generate a random domain-specific target style as $\widetilde{\mathbf{s}} = {M}(\mathbf{w},\widetilde{y}_{tar})$ from a random domain-independent style $\mathbf{w}$ and random target domain $\widetilde{y}_{tar} \in \mathcal{Y}_{tar}$. The decoder then synthesizes the translated image as $G(\hat{\mathbf{z}}, \widetilde{\mathbf{s}})$. Finally, we employ adversarial loss~\cite{goodfellow2014generative} as
\begin{equation}
\begin{split}
	\mathcal{L}_{\text{adv}} = & \mathbb{E}_{\mathbf{x}, y_{src}} \left[ \log{D^{T}\left(\mathbf{x},y_{src}\right)} \right] \\
	& + \mathbb{E}_{\mathbf{x}, \widetilde{y}_{tar}, \mathbf{w}}\left[\log{\left(1 - D^{T}\left(G\left(\hat{\mathbf{z}}, \widetilde{\mathbf{s}}\right),\widetilde{y}_{tar}\right)\right)}\right],
	\label{eqn:adv_loss}
\end{split}
\end{equation}
where $G$ tries to generate images $G(\hat{\mathbf{z}}, \widetilde{\mathbf{s}})$ that look similar to images from domain $\widetilde{y}_{tar}$, while the discriminator $D^{T}$ aims to distinguish between translated samples $G(\hat{\mathbf{z}}, \widetilde{\mathbf{s}})$ and real samples $\mathbf{x}$ from domain $y_{src}$. 
$G$ aims to minimize this objective against an adversary $D^{T}$ that tries to maximize it, i.e.,  $\min_G \max_{D^T} \mathcal{L}_{\text{adv}}(G,D^{T},\mathcal{Y}_{src},\mathcal{Y}_{tar}).$

\minisection{Style reconstruction.} We encourage the decoder to optimize the style representation $\widetilde{\mathbf{s}}$ when generating the image $G(\hat{\mathbf{z}},\widetilde{\mathbf{s}})$ with a style reconstruction loss 
\begin{equation}
	\mathcal{L}_{\text{sty}} = \mathbb{E}_{\mathbf{x}, \widetilde{y}_{tar}, \mathbf{w}} \left[{\left\|\widetilde{\mathbf{s}} - {E^{s}}\left(G\left(\hat{\mathbf{z}},\widetilde{\mathbf{s}}\right),\widetilde{y}_{tar}\right)\right\|}_{1} \right].
	\label{eqn:sty_rec_loss}
\end{equation}

\minisection{Style diversification.} To encourage diversity and prevent mode collapse, we sample and map random pairs of styles $\widetilde{\mathbf{s}}_{1}=M\left(\mathbf{w}_{1},\widetilde{y}_{tar}\right)$ and $ \widetilde{\mathbf{s}}_{2}=M\left(\mathbf{w}_{2},\widetilde{y}_{tar}\right)$, using diversity sensitive loss
\begin{equation}
	\mathcal{L}_{\text{ds}} = \mathbb{E}_{\mathbf{x}, \widetilde{y}_{tar}, \mathbf{w_{1}}, \mathbf{w_{2}}} \left[ { \lVert G\left(\hat{\mathbf{z}},\widetilde{\mathbf{s}}_1\right) - G\left(\hat{\mathbf{z}},\widetilde{\mathbf{s}}_2\right) \lVert}_{1} \right].
	\label{eqn:ds_loss}
\end{equation}

\minisection{Cycle consistency.} To ensure that the domain-invariant structure of the input image $\mathbf{x}$ in the translated image $G\left(\mathbf{z}, \widetilde{s}\right)$ is preserved we use the cycle consistency mechanism~\cite{zhu2017unpaired}
\begin{equation}
	\mathcal{L}_{\text{cyc}} = \mathbb{E}_{\mathbf{x}, y_{src}, \widetilde{y}_{tar}, \mathbf{w}} \left[ {\lVert\mathbf{x} - G\left(E^{c}\left(G\left(\hat{\mathbf{z}}, \widetilde{\mathbf{s}}\right)\right), \hat{\mathbf{s}}\right)\rVert}_{1} \right],
\end{equation}
where $\hat{\mathbf{s}}=E^{s}(\mathbf{x}, y_{src})$ is the style of the input image.

\minisection{Rate.} We estimate the rate as the entropy of the bitstream via modeling the distribution of $\mathbf{z}$ using the entropy model $P$. This term encourages the model to retain the most important information in a compact representation
\begin{equation}
	\mathcal{L}_{\text{rate}} = \mathbb{E}_\mathbf{x}\left[-\log\left(P\left(\mathbf{z}\right)\right)\right]
\end{equation}
using $\mathbf{z}=E^{c}\left(\mathbf{x}\right)$.
The final loss is
\begin{equation}
\begin{split}
	\mathcal{L}_{T} = &\mathcal{L}_{\text{adv}} + {\gamma}_{\text{sty}} \thinspace \mathcal{L}_{\text{sty}} - {\gamma}_{\text{ds}}\thinspace\mathcal{L}_{\text{ds}} + {\gamma}_{\text{cyc}}\thinspace\mathcal{L}_{\text{cyc}} \\
	& + {\lambda}_{T}\thinspace\mathcal{L}_{\text{rate}}.
\end{split}
\label{eq:I2Icodec_loss}
\end{equation}

\section{Unified translation and autoencoding}
While the I2Icodec framework can realize distributed I2I translation, being able to recover the original input image (i.e. regular image compression) is equally important in practice. In order to avoid having to deploy two independent codecs (i.e. I2Icodec and autoencoding codec), here we propose a unified framework to transmit an input image and recover either  reconstruction of the original image (autoencoding mode) or a translated image (translation mode), in a single model. 

\subsection{UI2Icodec framework}
As shown in Figure~\ref{model}, we endow the I2Icodec with a switching mechanism controlled via the mode input $m\in\left\{\text{A},\text{T}\right\}$, which signals the operating mode. In the following, we describe the additional modifications to the I2Icodec framework to implement the joint functionality.

\minisection{Adaptive encoder and decoder.} The content encoder $A\_E^{c}\left(\mathbf{x};m\right)$ and the decoder $A\_G\left(\hat{\mathbf{z}},\mathbf{s};m\right)$ are conditioned on the mode $m$. When $m=T$, the encoder and decoder operate exactly as the I2Icodec described earlier; When $m=A$, the model works in autoencoding mode, i.e. normal neural image compression. 

\begin{figure*}[!t]
	\centering
	\subfigure[AdaResBlock in adaptive encoder $A\_E^{c}$]{\label{fig:a}\includegraphics[width=62mm]{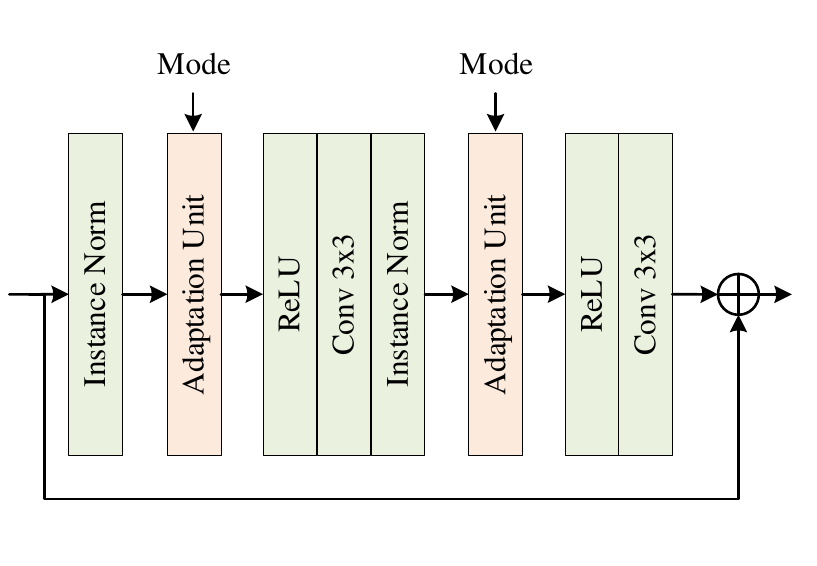}}
	\subfigure[AdaResBlock in adaptive decoder $A\_G$]{\label{fig:b}\includegraphics[width=62mm]{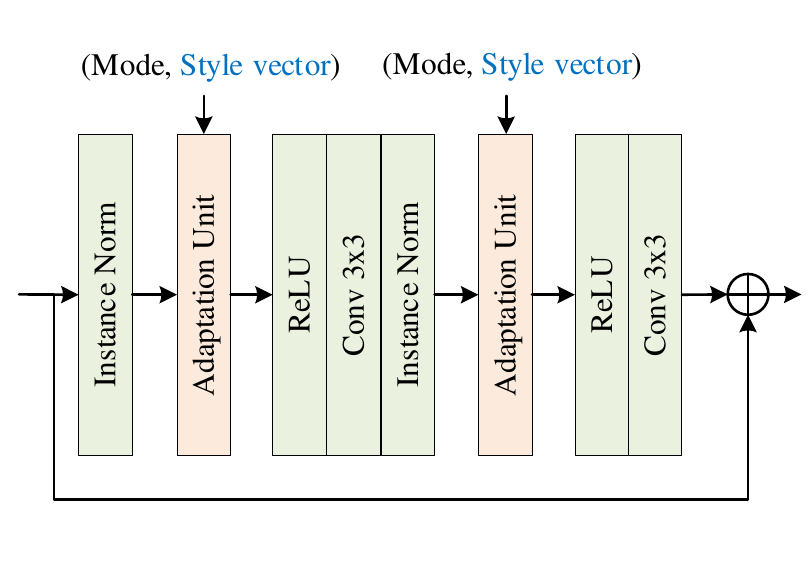}}
	\subfigure[Adaptation units in $A\_E^{c}$ and $A\_G$]{\label{fig:c}\includegraphics[width=46mm]{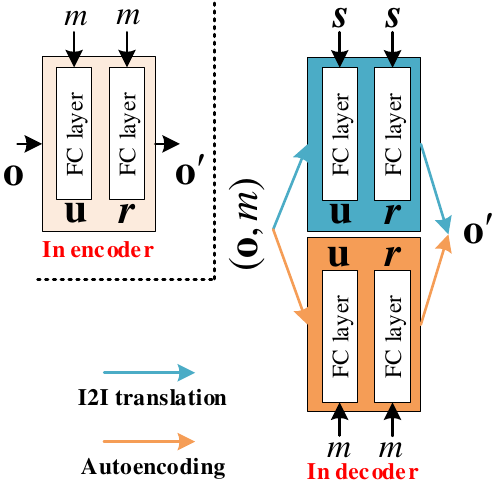}}
	\caption{(a-b): Adaptive ResBlock for switching between I2I translation and autoencoding; (c) Details of adaptation units in encoder $A\_E^{c}$ and decoder $A\_G$.}
	\label{fig:adaresblocks}
\end{figure*}

\begin{figure*}[!h]
  \centering
  \includegraphics[width=0.98\textwidth]{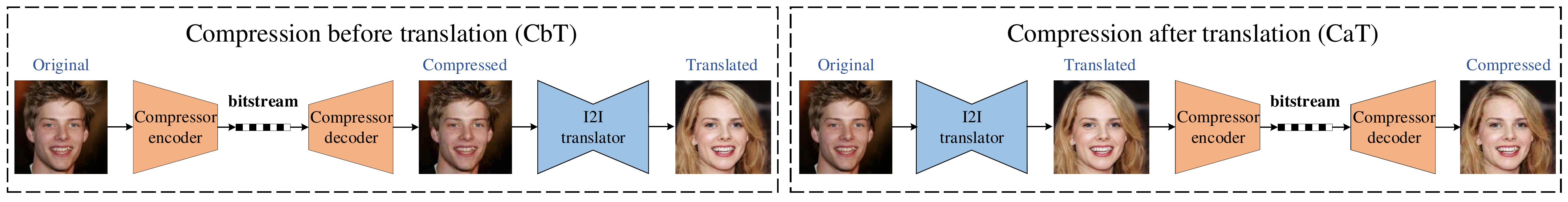}
  \caption{Baselines for distributed I2I translation.}
  \label{CATCBT}
\end{figure*}

\minisection{Adaptive residual blocks (AdaResBlocks).} The switching functionality is implemented by conditioning the residual blocks of the content encoder and decoder on the mode $m$, via an adaptation unit that modulates intermediate features within the residual block (see Figure~\ref{fig:adaresblocks} (a) and (b)). As shown in Figure~\ref{fig:adaresblocks} (c), adaptation units of the content encoder modulate a given input feature $\mathbf{o}$ as $\mathbf{o}'=\mathbf{u}\left(m\right) \odot \mathbf{o}+\mathbf{r}\left(m\right)$. The adaptation units in the decoder implement   $\mathbf{o}'=\mathbf{u}\left(\mathbf{s};m\right) \odot \mathbf{o}+\mathbf{r}\left(\mathbf{s};m\right)$. In both cases,  scale and bias parameters themselves are obtained via linear functions of $m$ (and $\mathbf{s}$ in the decoder).

\minisection{Conditional entropy model.} We condition the hyperprior~\cite{balle2018variational} on the mode, i.e. $C\_P(\mathbf{z};m)$, and one underlying factorized model for translation and another for autoencoding, selected depending on $m$.

\minisection{Task-specific discriminators.} We use a separate discriminator $D^{A}$ (see Figure~\ref{model}) when optimizing the autoencoding task. We found this to be more effective than using a shared discriminator for both tasks.

\subsection{Losses}
During training we optimize a loss with two terms corresponding to each of the operating modes 
\begin{equation}
	\label{eq:total_loss}
	\mathcal{L} = \left[m=\text{A}\right]\mathcal{L}_{\text{A}} + \left[m=\text{T}\right]\mathcal{L}_{\text{T}},
\end{equation}
where $\left[I\right]$ is the Iverson bracket (1 when $I$ is true, 0 otherwise), $\mathcal{L}_{\text{T}}$ is the loss described in the previous section minimized when $m=\text{T}$, and $\mathcal{L}_{\text{A}}$ is the autoencoding loss, minimized for $m=\text{A}$. The latter combines the losses of rate-distortion tradeoff and GAN loss as $\mathcal{L}_{T} = \mathcal{L}_{\text{RD}} + \beta \mathcal{L}_{\text{adv2}}$ , which are introduced in the following. 

\minisection{Rate-distortion}. In autoencoding we optimize a combination of rate and distortion, where the tradeoff is controlled by the parameter $\lambda_A$ which also decides the final bit-rate of compression model. The loss is
\begin{equation}
\mathcal{L}_{\text{RD}}=\mathbb{E}_{\mathbf{x}}\left[d\left(\mathbf{x}, \hat{\mathbf{x}}\right)-\lambda_{A} \log\left(C\_P\left(\mathbf{z}; m=A\right)\right) \right]
\end{equation}
where $\mathbf{z}=A\_E^{c}(\mathbf{x}; m=\text{A})$ is the latent representation, $\hat{\mathbf{x}}=A\_G(\hat{\mathbf{z}};m=\text{A})$ is the reconstructed image, and  $d\left(\mathbf{x}, \hat{\mathbf{x}}\right)$ is the distortion metric (mean-squared error in our case). 

\minisection{Adversarial loss for generative image compression.} We also encourage realism in the reconstructed images by introducing adversarial training with the discriminator $D^{A}$, which has been verified that it can help to generate high fidelity images even with an extreme low rate~\cite{mentzer2020high}. Similarly to $\mathcal{L}_{\text{adv}}$, we apply the second adversarial loss as 
\begin{equation}
\begin{split}
	\mathcal{L}_{\text{adv2}} = &\mathbb{E}_{(\mathbf{x},y_{src})} \left[ \log{D^A\left(\mathbf{x},y_{src}\right)} \right] \\
	&+ \mathbb{E}_{(\mathbf{x},y_{src})}\left[\log{\left(1 - D^A\left(\hat{\mathbf{x}},y_{src}\right)\right)}\right].
	\label{eqn:adv_loss2}
\end{split}
\end{equation}
Then, optimizing $\min_{A\_G} \max_{D^A} \mathcal{L}_{\text{adv2}}(A\_G,D^{T},\mathcal{Y}_{src},\mathcal{Y}_{src})$ will lead $A\_G$ with $m=A$ to generate more realistic images.

\begin{figure*}[!h]
  \centering
  \subfigure[FID on AFHQ]{\label{FID_afhq}\includegraphics[width=82mm]{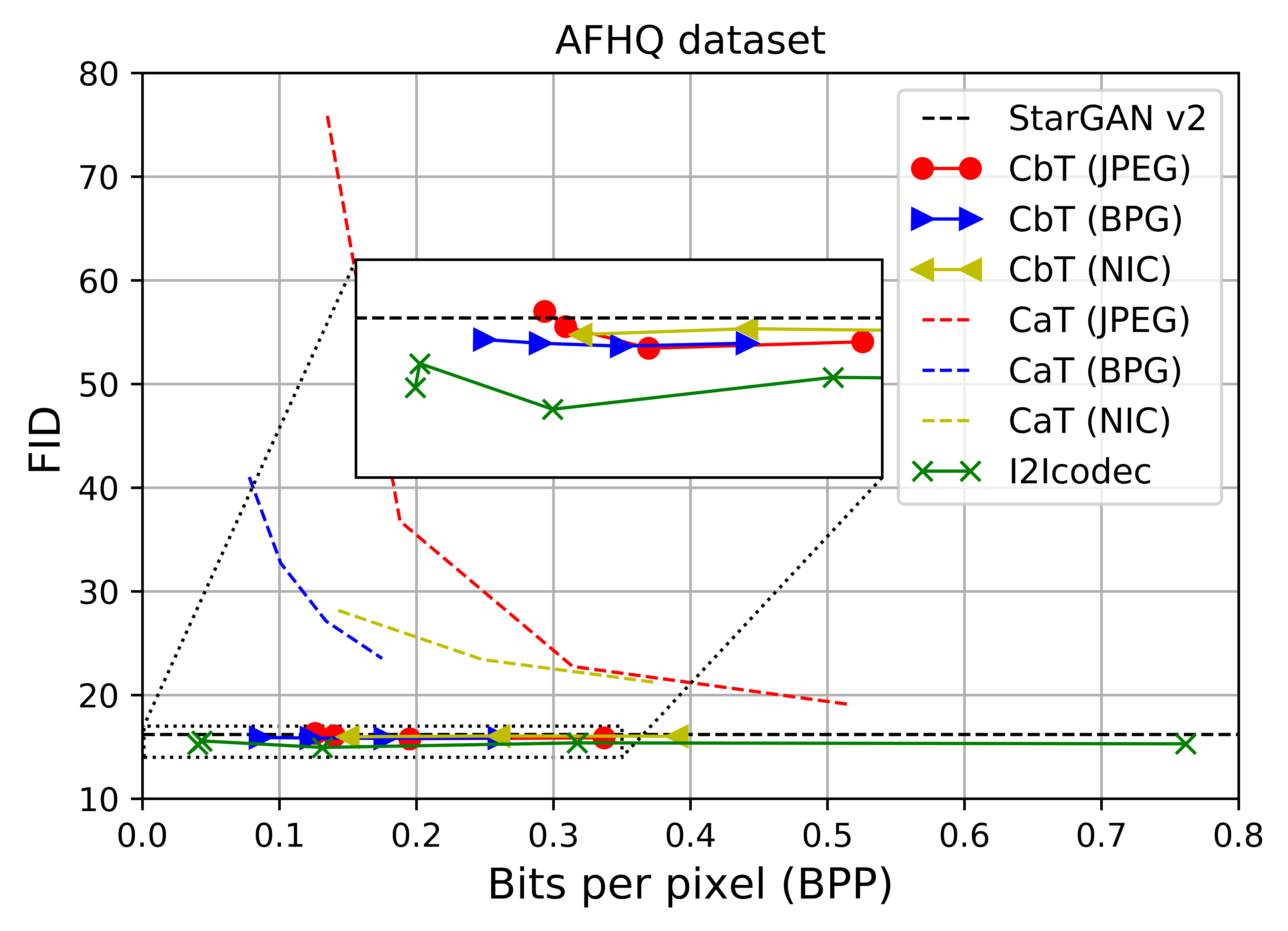}}
  \subfigure[LPIPS on AFHQ]{\label{LPIPS_afhq}\includegraphics[width=83mm]{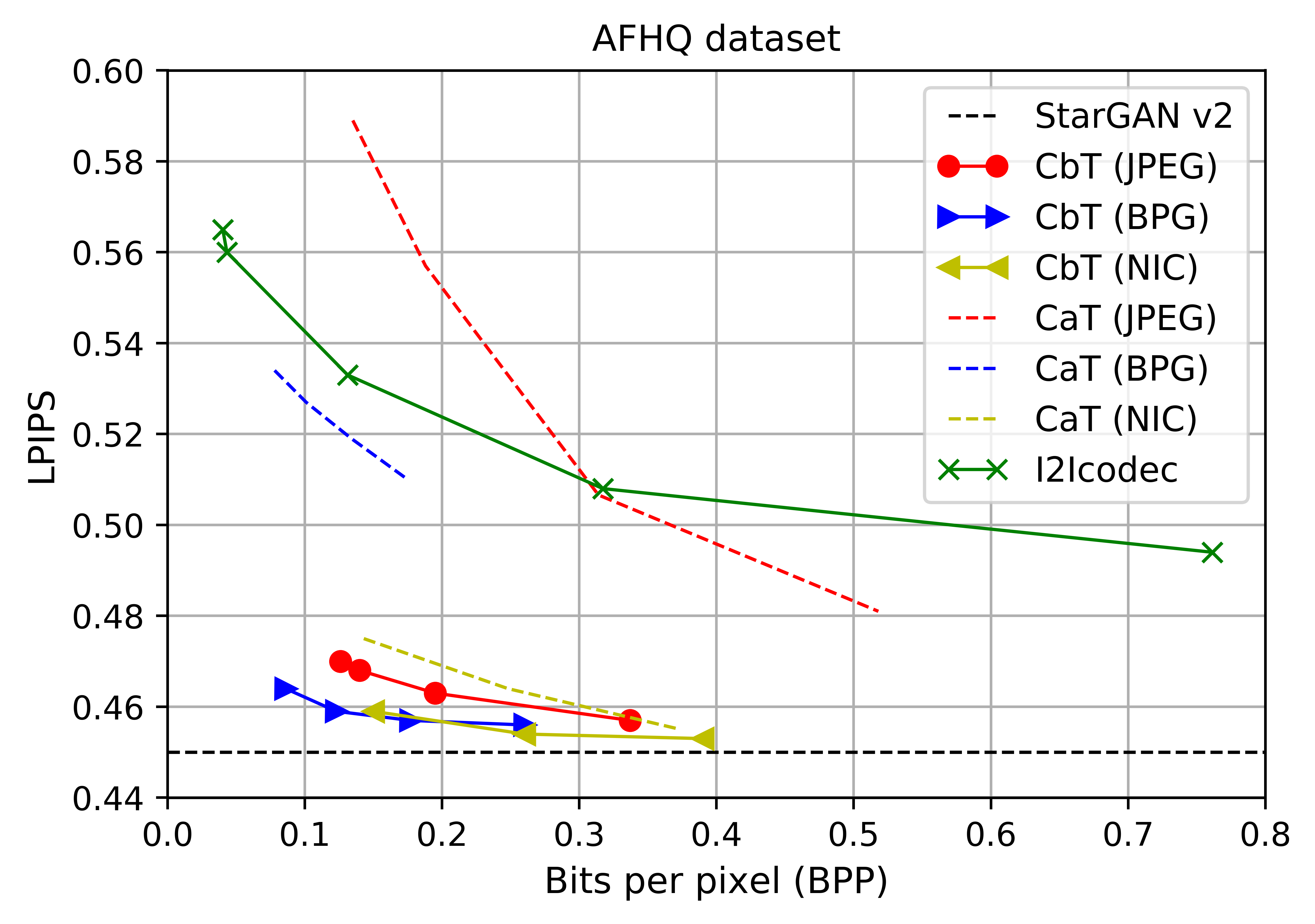}}
  \subfigure[FID on CelebA-HQ]{\label{FID_celeba}\includegraphics[width=82mm]{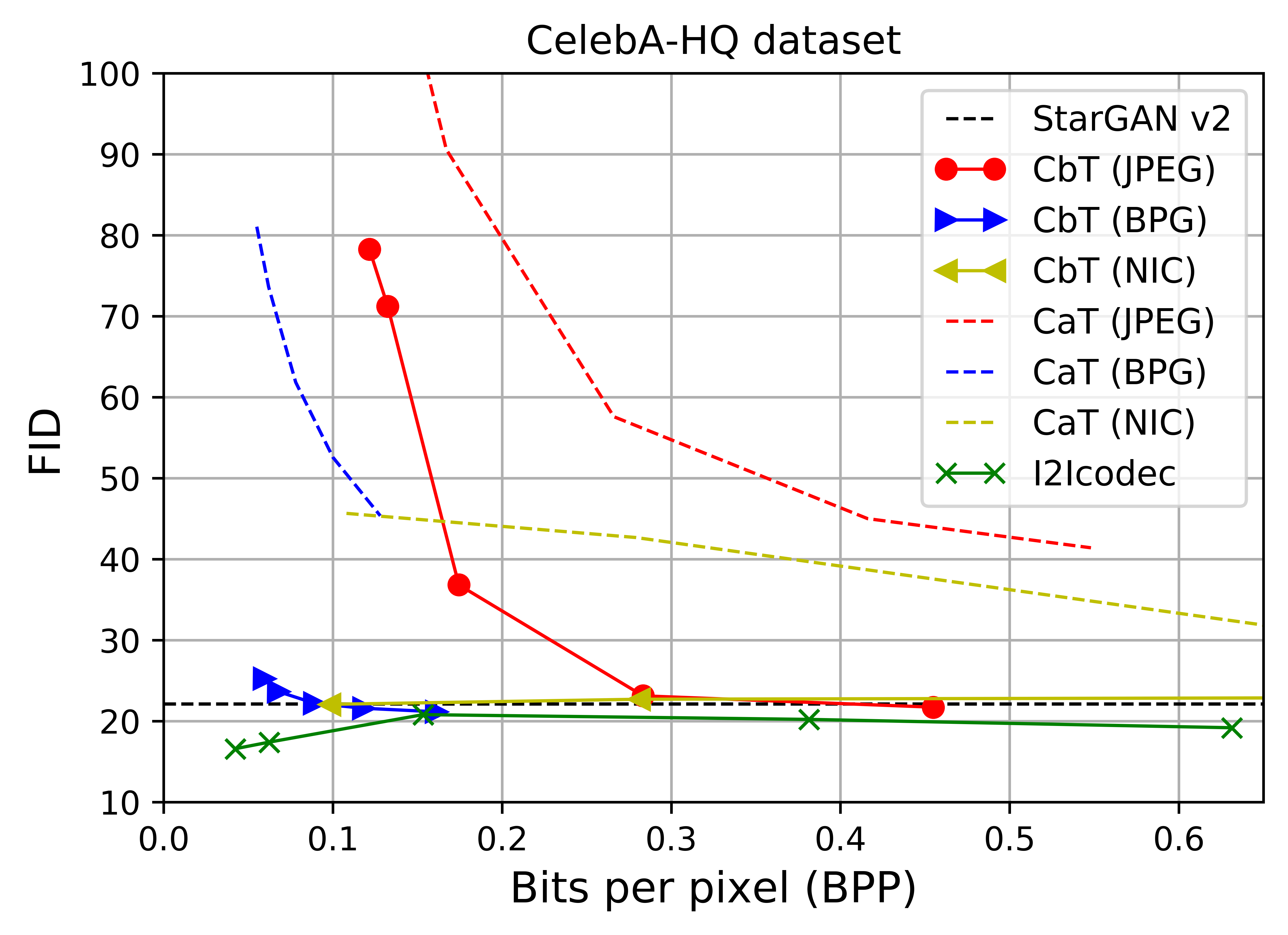}}
  \subfigure[LPIPS on Celeb-HQ]{\label{LPIPS_celeba}\includegraphics[width=83mm]{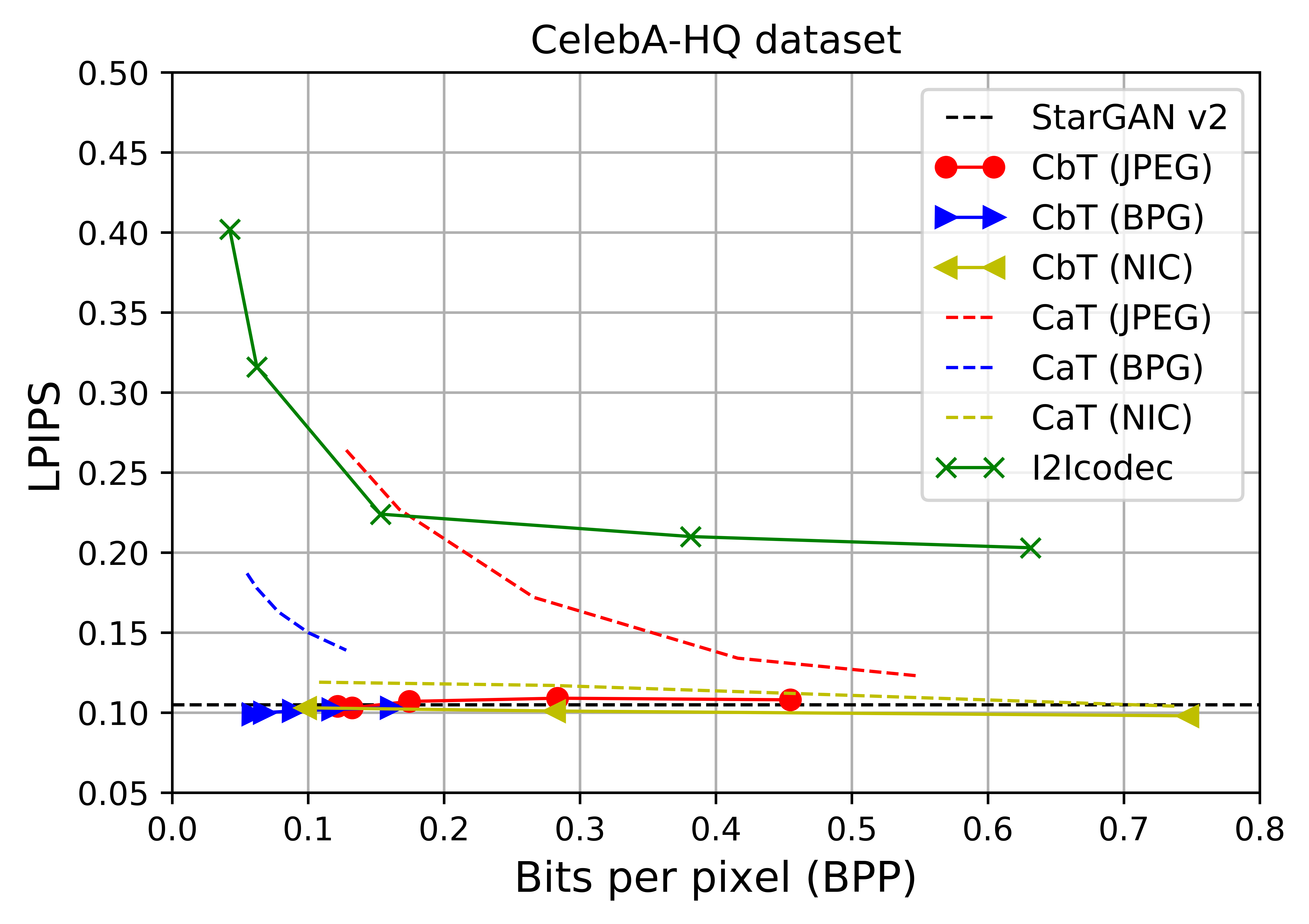}}
  \caption{Results of different scheme on AFHQ and CelebA-HQ dataset.}
  \label{comparison_afhp}
\end{figure*}

\begin{figure*}[!t]
  \centering
  \subfigure[AFHQ dataset.]{\label{I2Icodec_animal}\includegraphics[width=85mm]{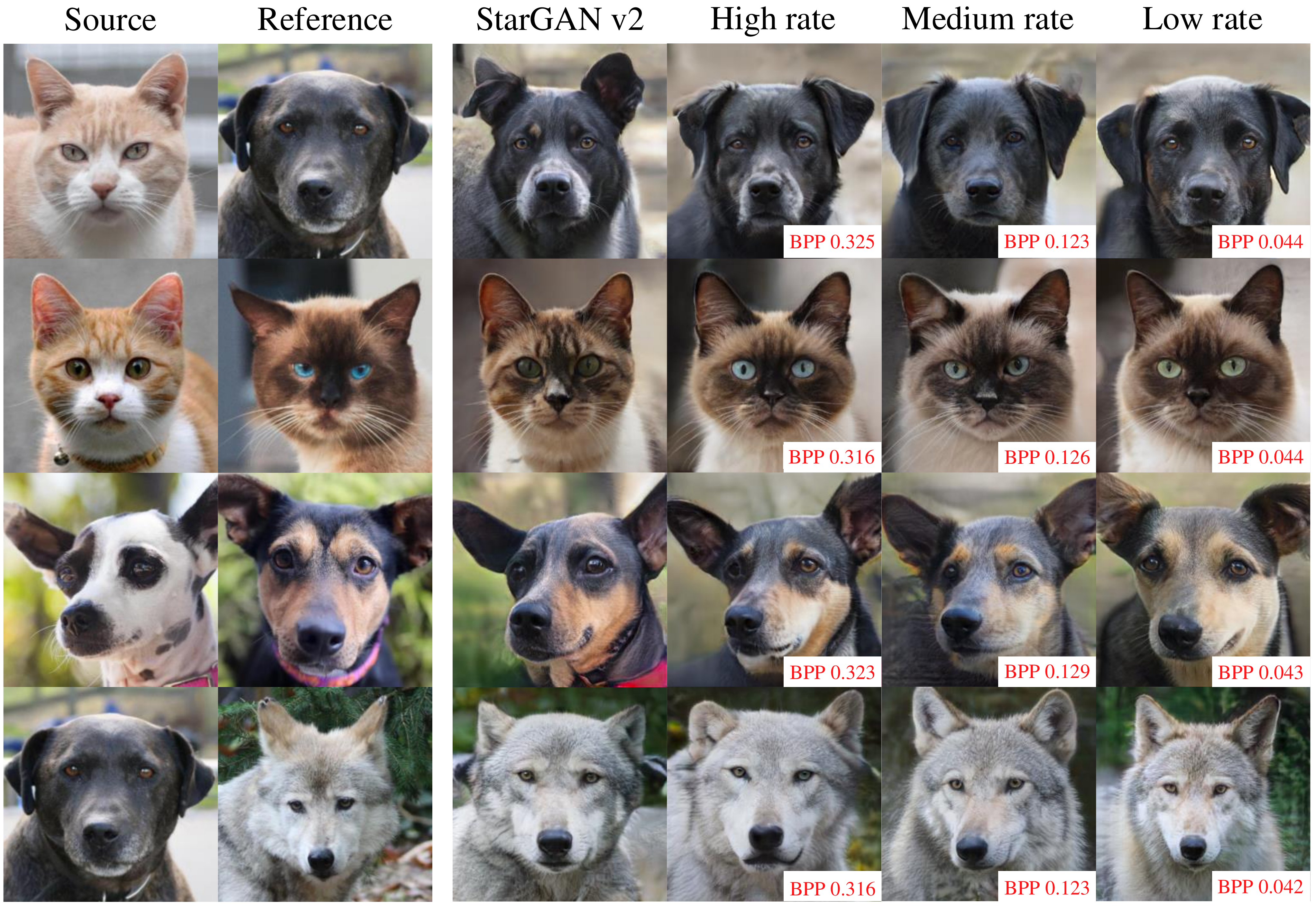}}
  \subfigure[CelebA-HQ dataset]{\label{I2Icodec_human}\includegraphics[width=85mm]{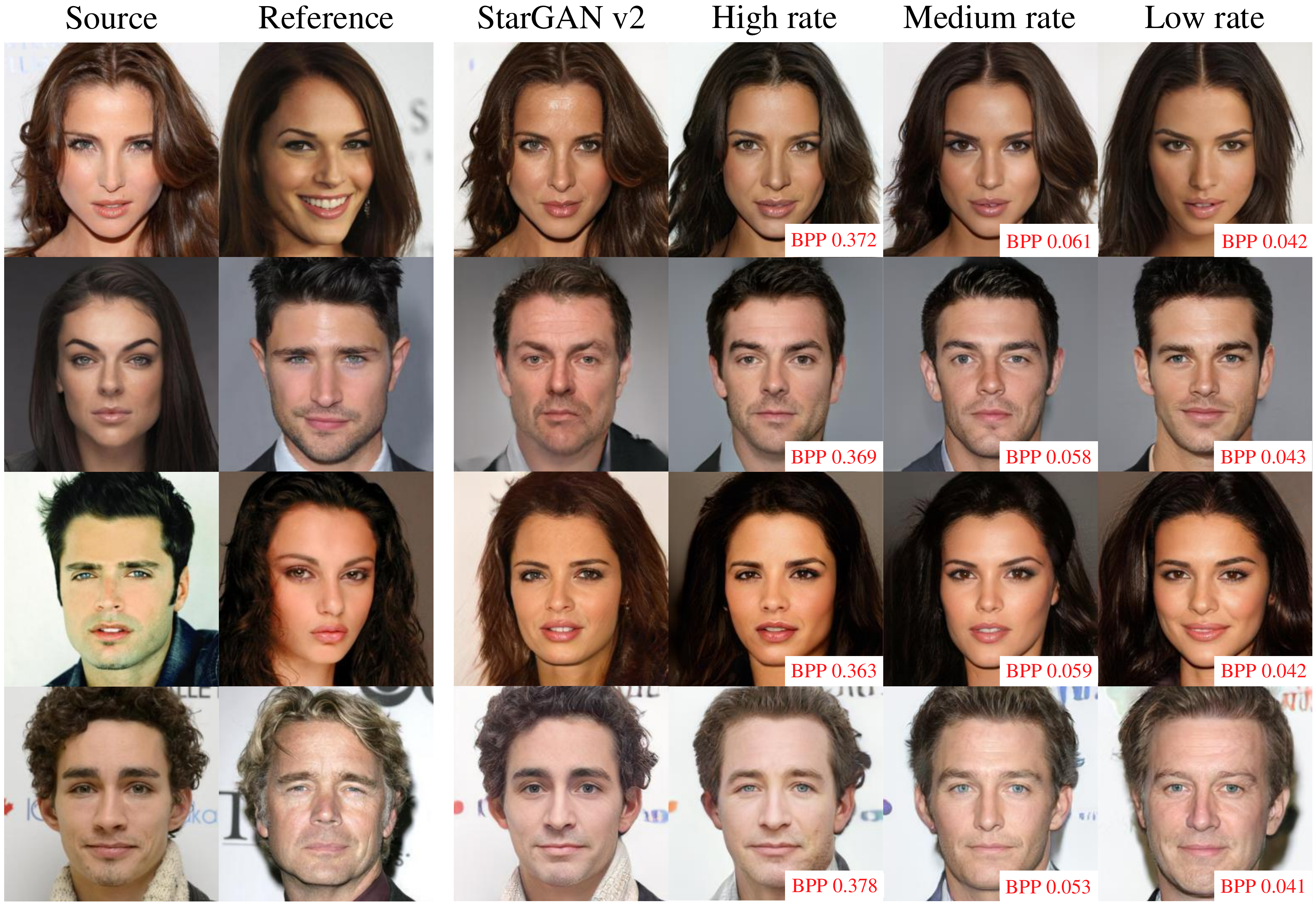}}
  \caption{Translated images with reference on different rate.}
  \label{I2Icodec_reference}
\end{figure*}

\begin{table*}[t]
\centering
\resizebox{0.99\textwidth}{!}{%
  \begin{tabular}{|c|c|c|c|c|c|c|c|c|c|c|c|c|}
    \hline
    \multirow{3}{*}{\textbf{Method}}&\multicolumn{6}{c|}{\textbf{Latent-guided synthesis}} &\multicolumn{6}{c|}{\textbf{Reference-guided synthesis}}\\
    \cline{2-13}
     &\multicolumn{3}{c|}{\textbf{CelebA-HQ}} &\multicolumn{3}{c|}{\textbf{AFHQ}} &\multicolumn{3}{c|}{\textbf{CelebA-HQ}} &\multicolumn{3}{c|}{\textbf{AFHQ}}\\
    \cline{2-13}
     &\textbf{FID}$\downarrow$ &\textbf{LPIPS} $\uparrow$ &\textbf{BPP} &\textbf{FID}$\downarrow$  &\textbf{LPIPS}$\uparrow$ &\textbf{BPP} &\textbf{FID}$\downarrow$  &\textbf{LPIPS}$\uparrow$ &\textbf{BPP} &\textbf{FID} $\downarrow$ &\textbf{LPIPS} $\uparrow$ &\textbf{BPP}\\      
    \hline
    \hline
    MUNIT~\cite{huang2018multimodal} &31.4 &0.363 &- &41.5 &0.511 &- &107.1 &0.176 &- &223.9 &0.199 &-\\
    DRIT~\cite{Lee2018drit} &52.1 &0.178 &- &95.6 &0.326 &- &53.3 &0.311 &- &114.8 &0.156 &-\\
    MSGAN~\cite{mao2019mode} &33.1 &0.389 &- &61.4 &0.517 &- &39.6 &0.312 &- &69.8 &0.375 &-\\
    StarGANv2~\cite{choi2020stargan}&22.1* &0.115* &- &16.2 &0.450 &- &23.3* &0.209* &- &19.8 &0.432 &-\\
    \hline
    I2Icodec ($\lambda_{T} = 0.5$) &16.6 &0.402 &0.043 &15.2 &0.565 &0.042  &21.0 &0.354 &0.043 &20.6 &0.526 &0.042\\
    I2Icodec ($\lambda_{T} = 0.1$) &20.8 &0.224 &0.153 &14.9 &0.533 &0.132 &20.7 &0.247 &0.153 &20.0 &0.494 &0.132\\
    I2Icodec ($\lambda_{T} = 0.05$) &20.2 &0.210 &0.381  &15.4 &0.508 &0.317 &20.0 &0.220 &0.381 &19.9 &0.471 &0.317\\
    \hline
    T + A (w GAN) &20.0 &0.088 &- &25.8 &0.288 &- &22.2 &0.082 &- &28.1 &0.265 &-\\
    T + A (w/o GAN) &20.6 &0.098 &- &28.7 &0.268 &- &21.1 &0.089 &- &32.5 &0.238 &-\\
    \hline
    UI2Icodec (T mode) &18.14 &0.403 &0.065 &13.5 &0.531 &0.131 &17.45 &0.360 &0.065 &17.9 &0.496 &0.131\\
    \hline
    \hline
    Real images &14.8 & - &- &12.9 & - &- &14.8 & - &- &12.9 & - &-\\
    \hline
    \end{tabular}}
    \caption{Quantitative comparison. 
The FIDs of real images are computed between the training and test sets. Note that they may not be optimal values since the number of test images is insufficient, but we report them for reference. * means the results of StarGAN v2 on CelebA-HQ are from the same model architecture on AFHQ, which doesn't include skip connections with the adaptive wing based heatmap~\cite{wang2019adaptive}.}
\label{I2I_performance}
\end{table*}

\section{Experiments}
In this section, we describe our experimental setup and results. We analyze the effects of the rate constraint of I2Icodec in the translations (see Section~\ref{sec:I2Icodec}). We also show the results of our unified framework UI2Icodec on both autoencoding and I2I translation (Section~\ref{sec:UI2Icodec}). Ablation study and additional results are shown in Section~\ref{sec:ablation}.

\textbf{Datasets.} Our experiments are mainly conducted on CelebA-HQ and the animal faces (AFHQ) datasets~\cite{choi2020stargan}. As in~\cite{choi2020stargan}, CelebA-HQ is separated into male and female domains, and AFHQ into cat, dog and wildlife domains. We resized all images to $256\times256$ for training and comparisons. 

\textbf{Training.} For I2Icodec, we train the model minimizing  $\mathcal{L}_{T}$ during 100k iterations. We set ${\gamma}_{sty}={\gamma}_{ds}={\gamma}_{cyc}=1$ which are same with ~\cite{choi2020stargan}, and $\lambda_{T} \in \left\{0.01, 0.05, 0.1, 0.3, 0.5\right\}$ (in Eq.~\ref{eq:I2Icodec_loss}) which can lead to different bit-rates for the distributed I2I translation. For UI2Icodec, we first train $A\_E^c$ and $A\_G$ just with distortion loss (mean square error) on autoencoding mode for 50k iterations, then the whole model is jointly trained with another 100k iterations. We use Adam~\cite{kingma2014adam} to optimize the model with the loss of Eq.~\ref{eq:total_loss} calculated by setting $m=A$ for autoencoding and $m=T$ for I2I translation during each iteration. We set $\lambda_{A}=[5, 10, 15, 20, 30]$ and $\beta=1$ to achieve different bit-rates for normal image compression, and fix $\lambda_{T} = 0.5$ for translation on CelebA-HQ dataset and $\lambda_{T} = 0.1$ on AFHQ dataset.

\textbf{Evaluation metrics.} We rely on the usual metrics used in I2I translation and image compression. For translation we compute Fréchet inception distance (FID)~\cite{heusel2017gans} and learned perceptual image patch similarity (LPIPS)~\cite{zhang2018unreasonable} to evaluate quality and diversity, respectively. Autoencoding is evaluated by computing the distortion metrics PSNR, MS-SSIM, and LPIPS. Note that, for translation LPIPS is computed between pairs of translated images, while for autoencoding it is computed between original and reconstructed images. The rate is measured as the bits per pixel (BPP) of the bitstream including the content part output by $E^c$ of I2Icodec or $A\_E^c$ of UI2Icodec and the style part which is a 64-dimensional vector in Float32 and requires 0.03125 BPP for 256x256 images. Note that the contribution of other parts of the bitstream (i.e. operation mode) to the rate is negligible and ignored in our results.

\begin{figure*}[h]
  \centering
  \subfigure[AFHQ dataset.]{\label{I2Icodec_a}\includegraphics[width=86mm]{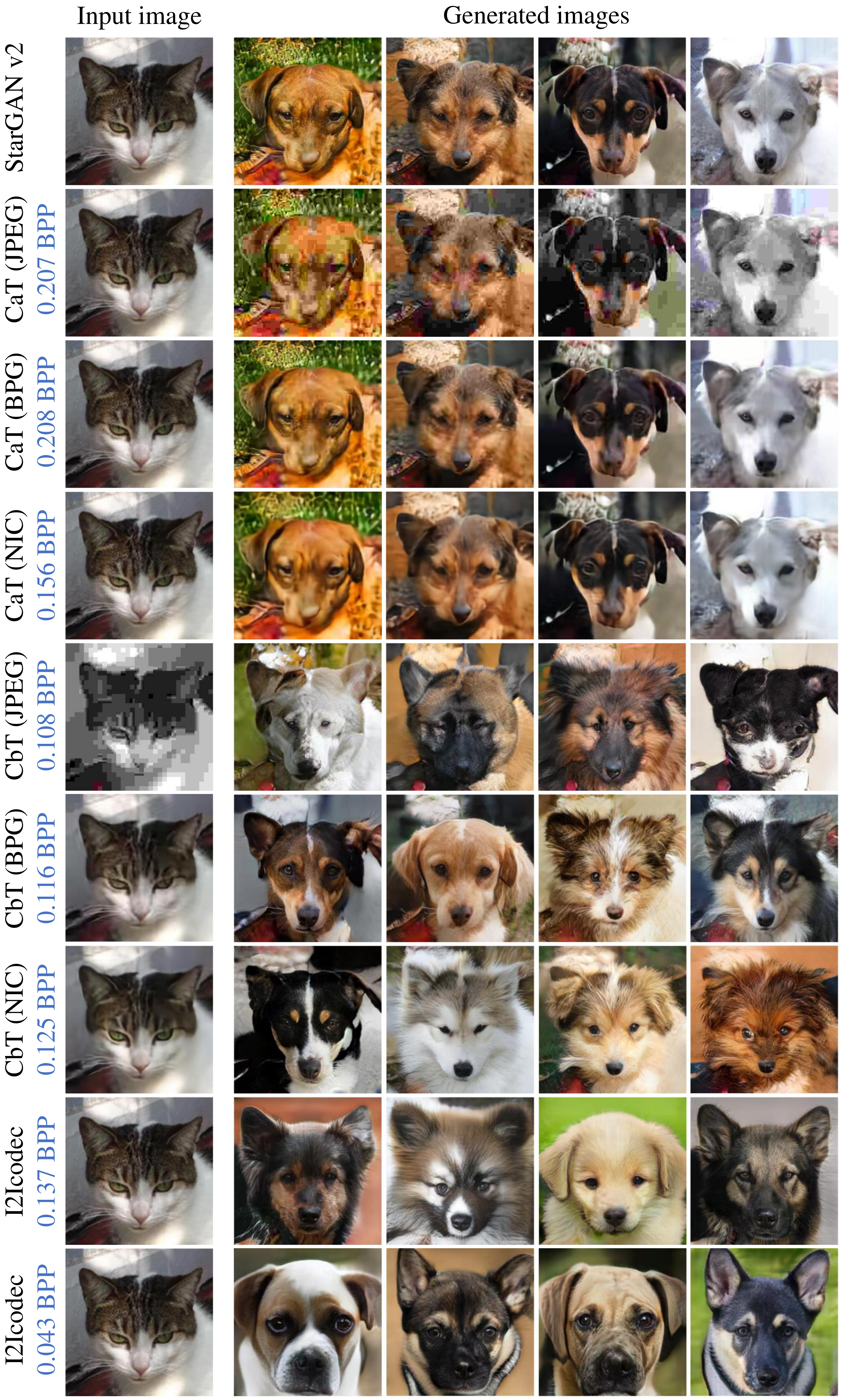}}
  \subfigure[CelebA-HQ dataset]{\label{I2Icodec_b}\includegraphics[width=86mm]{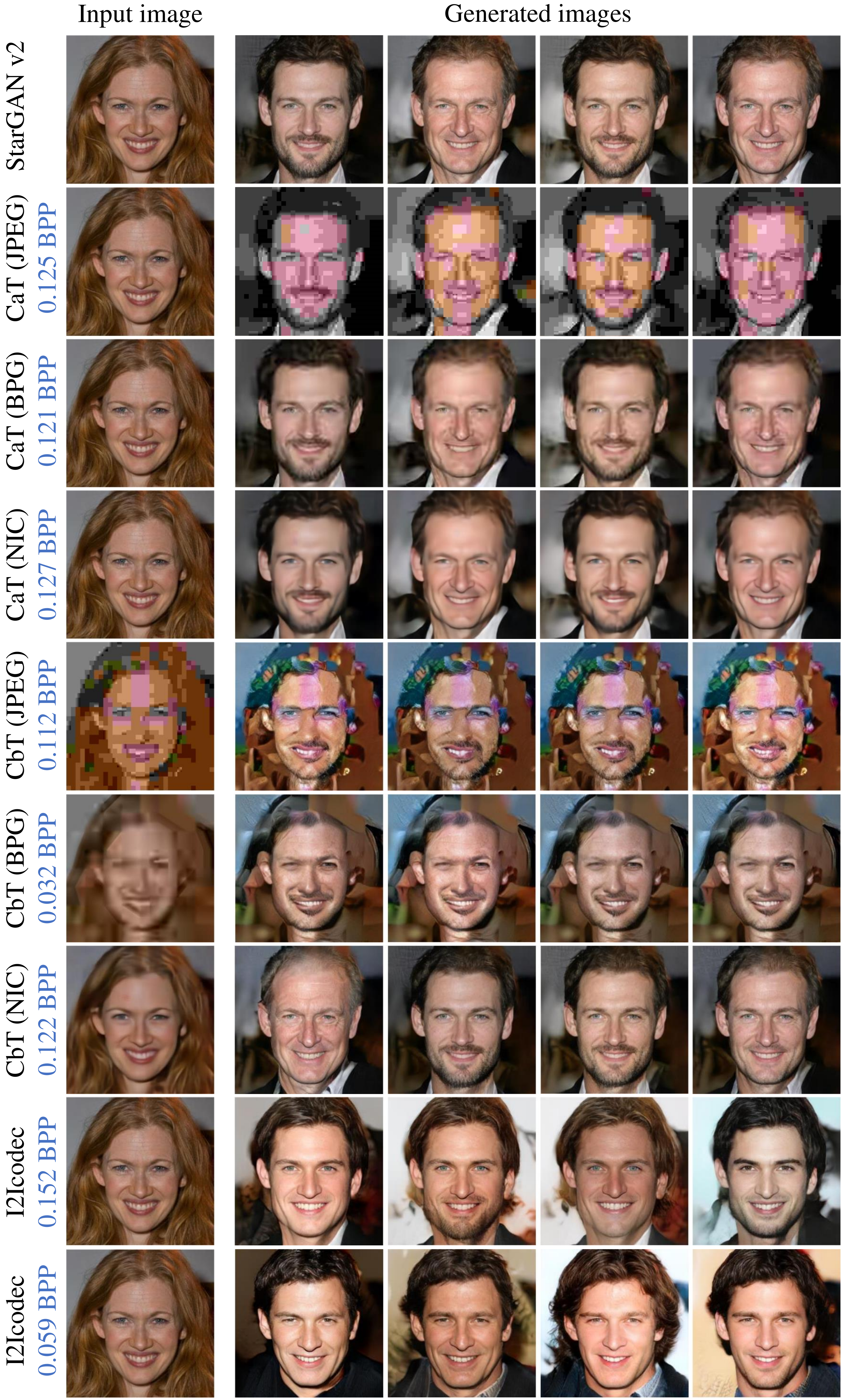}}
  \caption{Visualization of latent-guided synthesis with different methods (more samples in Section 5.3).}
  \label{I2Icodec_vis}
\end{figure*}

\subsection{Distributed I2I translation}
\label{sec:I2Icodec}

In this section, we analyze and compare different methods to address distributed I2I translation: (1) compression before translation (CbT): the input image is compressed and then translated after reconstruction, as shown in Figure~\ref{CATCBT} \textbf{left}; (2) compression after translation (CaT): the input image is translated and then compressed with the image codec, as shown in Figure~\ref{CATCBT} \textbf{right}; (3) the proposed I2Icodec (see Figure~\ref{model_I2I}). We use the pre-trained model same with ~\cite{choi2020stargan} as a translator, and two classic compression methods: JPEG and BPG as two compressor options for CbT and CaT. In addition, we also train domain-specific neural image compression models on CelebA-HQ and AFHQ separately (NIC in our experiments). We use the same encoder and decoder architecture of StarGAN v2 and a hyperprior entropy model~\cite{balle2018variational} for a fair comparison and optimized with mean square error. The training of NIC is the same with UI2Icodec but without translation mode. 

\textbf{Effect of compression on translation.} As shown in Figure~\ref{comparison_afhp}, we observe changes of both FID and LPIPS values with varying rates (BPP).
Notably, CbT always obtains lower FID scores than CaT, which is not surprising since CaT compresses translated images via one lossy codec and the final images have compression artifacts, resulting in worse FID. 
I2Icodec obtains the lowest FID among all methods at the same rate and also consists of only one encoder and decoder while CaT and CbT are not, which implies that I2Icodec needs less coding time. The diversity measured by LPIPS is shown in Figure~\ref{LPIPS_afhq} and Figure~\ref{LPIPS_celeba}, where CaT can achieve larger scores but with higher distortion, especially for CaT (JPEG) (see the second row in Figure~\ref{I2Icodec_vis}). CbT obtains a similar LPIPS as StarGAN v2, which is lower than our I2Icodec. In summary, I2Icodec achieves smaller bandwidth requirements for distributed I2I translation and provides an effective way to guide I2I translation by controlling the amount of information in the bottleneck via the rate constraint. In Table.~\ref{I2I_performance}, we also report the quantitative comparison with other I2I translation methods~\cite{huang2018multimodal, Lee2018drit, mao2019mode, choi2020stargan} that do not consider compression. It shows that I2Icodec can achieve a range of scores (FID and LPIPS) at different rates for both latent-guided and reference-guided synthesis on two datasets. We want to emphasize that I2Icodec is more efficient and effective than CaT and CbT and also provides a lever for I2I translation. 

\begin{table*}[t]
\centering
\resizebox{0.8\textwidth}{!}{%
  \begin{tabular}{|c|cccc|}
    \hline
    \textbf{Method} &NIC &StarGAN v2 &I2Icodec &UI2Icodec \\
    \hline
    \textbf{Number of parameters (millions)} &35.30 &53.73 &54.22 &54.23\\
    \hline
    \textbf{Training time (hours)} &10.3  &65.7 &71.34 &97.6 \\
    \hline
    \end{tabular}}
    \caption{Number of parameters and training time of different methods (CelebA-HQ dataset).}
\label{model_size_time}
\end{table*}

\begin{figure*}[t]
  \centering
  \includegraphics[width=0.98\textwidth]{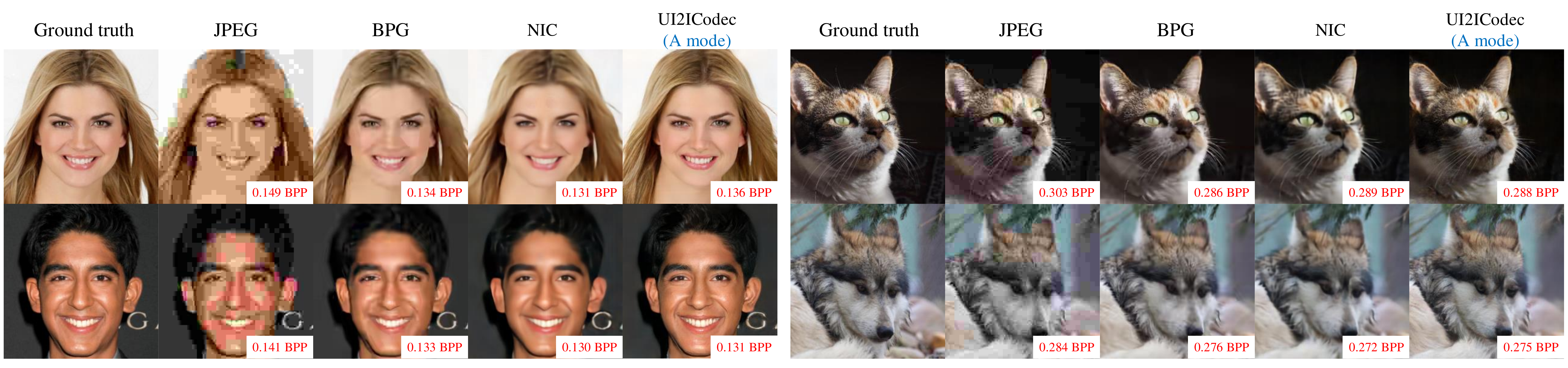}
  \caption{Reconstructions with different compression methods.}
  \label{rec_visualization}
\end{figure*}

\begin{figure*}[b]
  \centering
  \includegraphics[width=0.98\textwidth]{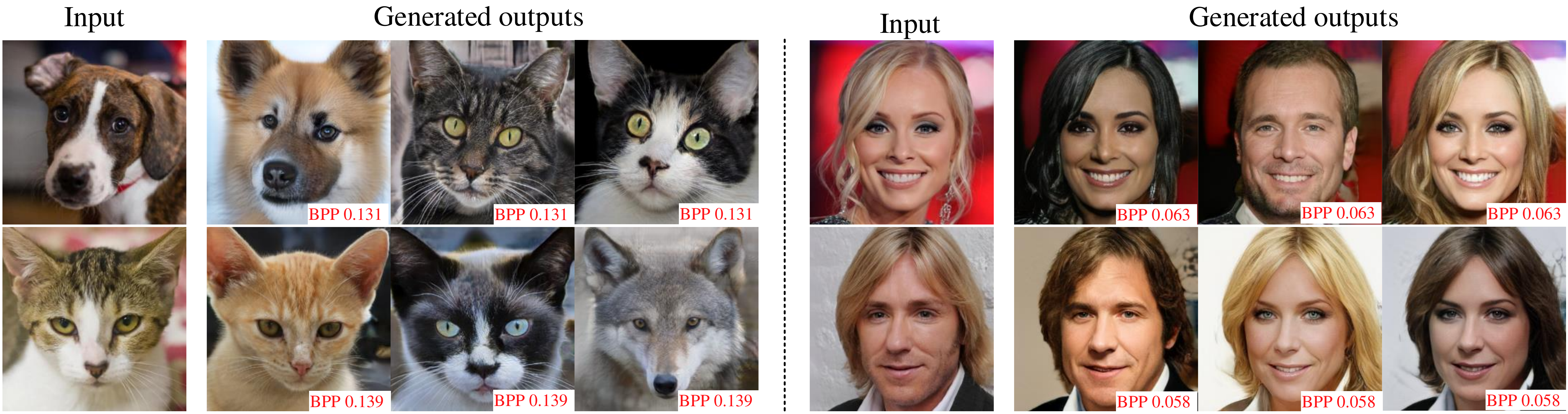}
  \caption{Diverse image synthesis results of UI2Icodec in the translation mode on the CelebA-HQ and AFHQ datasets.}
  \label{UI2Icodec_T_visualization}
\end{figure*}

\textbf{Visualization of translated images.} In Figure~\ref{I2Icodec_vis} we show translated images using different methods. It is obvious that CaT suffers from artifacts (see the second row) even with the better codec BPG (see the third row) and higher rate than other methods. While CaT with NIC as the compressor has less artifacts, the images are blurred (see the fourth row). In addition, different from than CaT with NIC and CbT with NIC, I2Icodec performs both compression and disentanglement jointly with transformation from pixel-level to latent space only once, which is a more efficient way. CbT with JPEG can keep some structure information, but result in unnatural translation (on AFHQ) or serious artifacts (on CelebA-HQ) due to the JPEG compression artifacts themselves. With BPG and NIC, the influence of compression is largely reduced (see the sixth and seventh rows in Figure~\ref{I2Icodec_a}), but note that it still appears again at low rates (see the sixth row in Figure~\ref{I2Icodec_b}). I2Icodec can generate more natural and diverse images even with extremely low rate.
In addition, we also show the synthesized images guided by a reference image on three different rates from high to low in Figure~\ref{I2Icodec_reference}. It shows that the translated images have a more similar style to reference image when the rate is lower, and also illustrate that this method can control well how much the translated image obtains the same style of reference image along with the rate.

\begin{figure*}[t]
  \centering
  \subfigure[PSNR $\uparrow$]{\label{RDPSNR_afhq}\includegraphics[width=57mm]{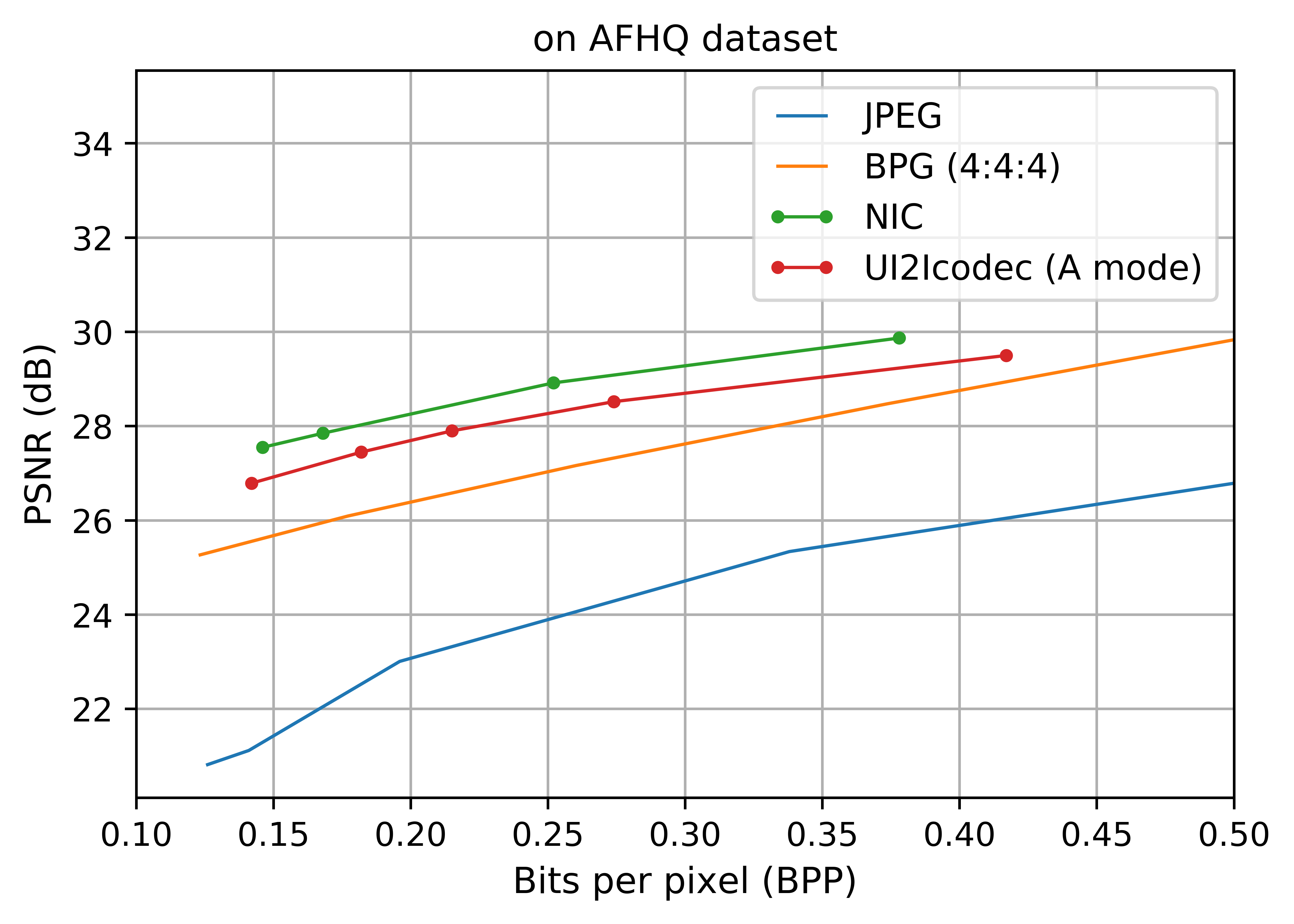}}
  \subfigure[MSSSIM $\uparrow$]{\label{RDMSSSIM_afhq}\includegraphics[width=57mm]{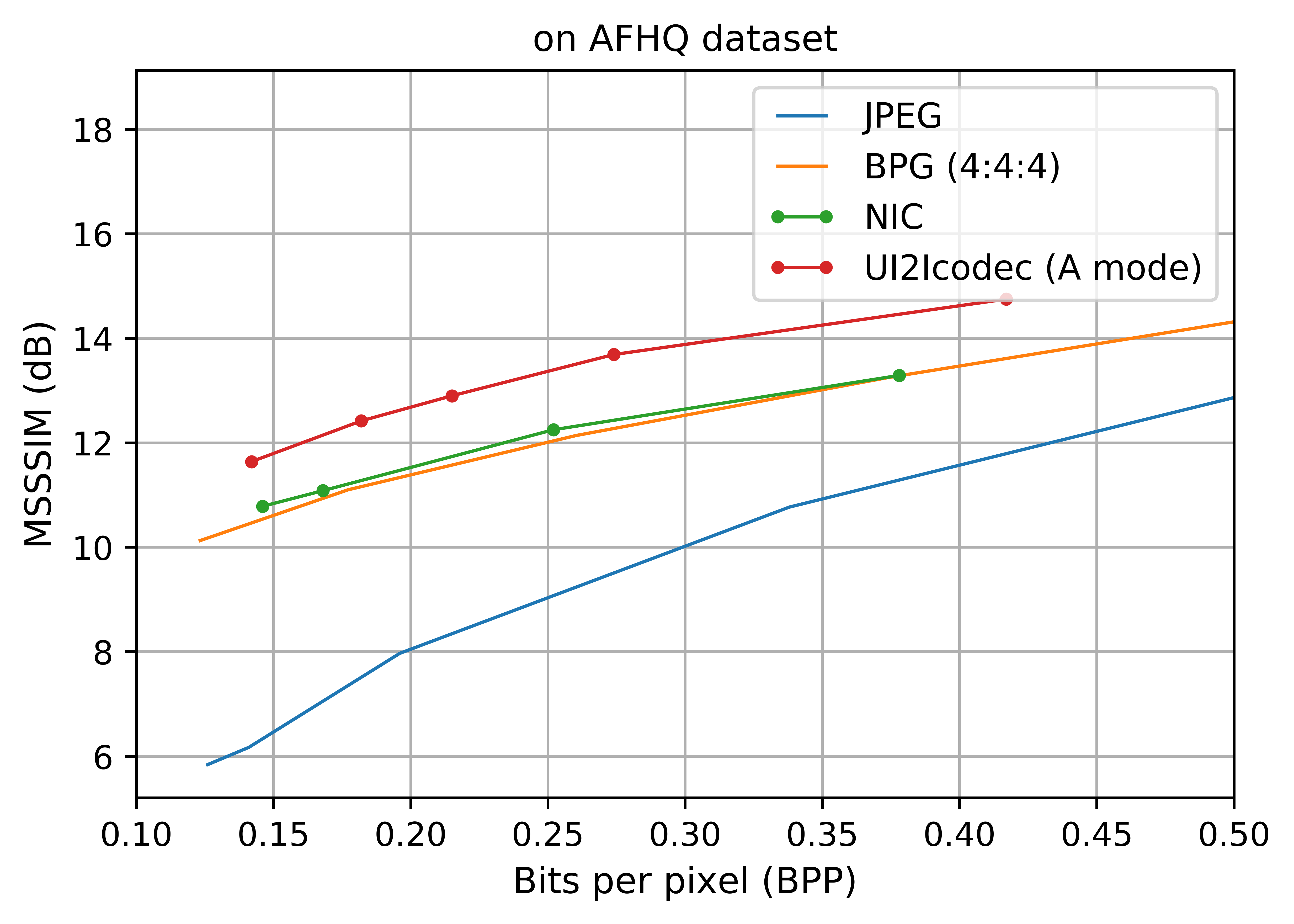}}
  \subfigure[LPIPS $\downarrow$]{\label{RDLPIPS_afhq}\includegraphics[width=57mm]{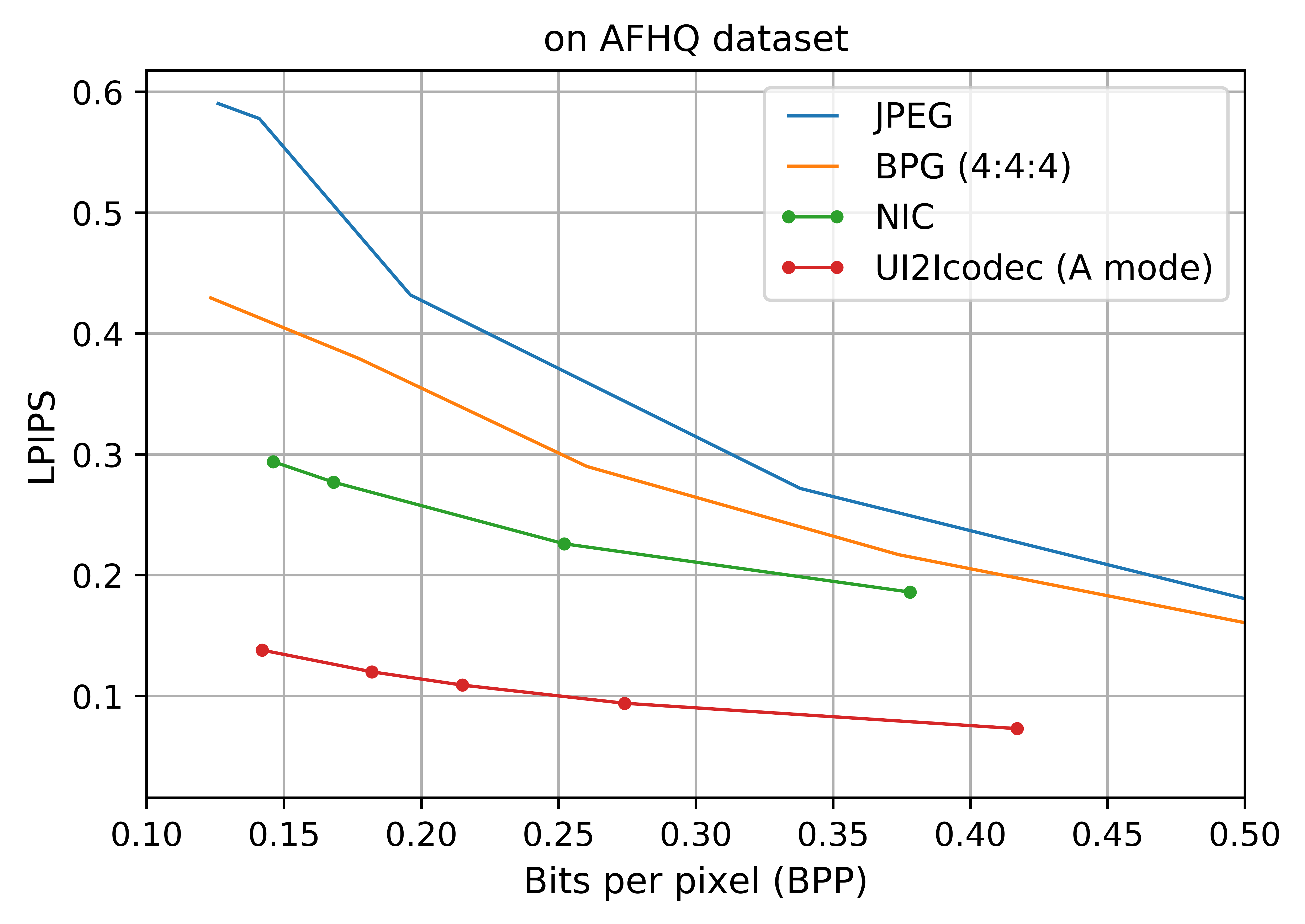}}
  \subfigure[PSNR $\uparrow$]{\label{RDPSNR_celebahq}\includegraphics[width=57mm]{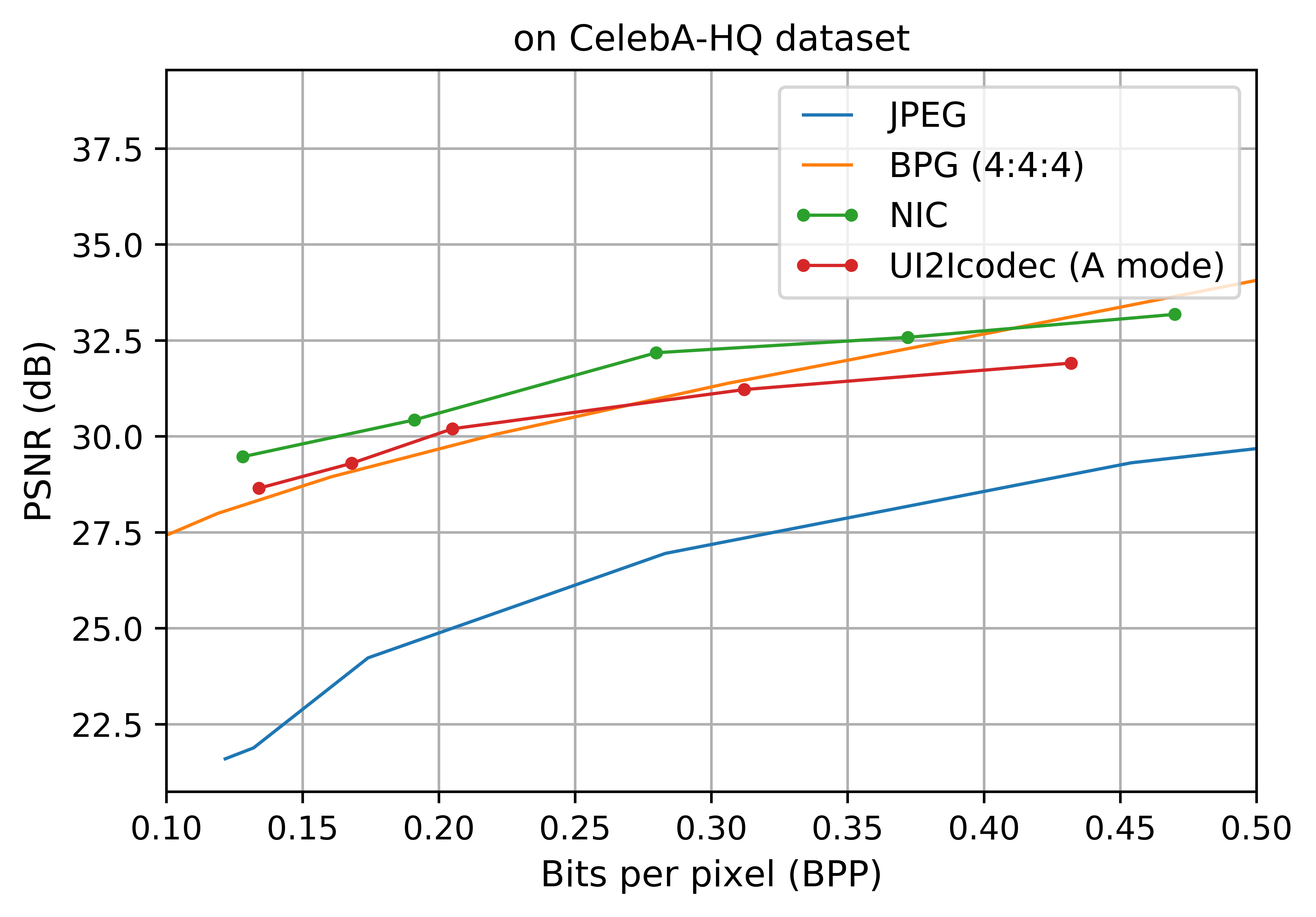}}
  \subfigure[MSSSIM $\uparrow$]{\label{RDMSSSIM_celebahq}\includegraphics[width=57mm]{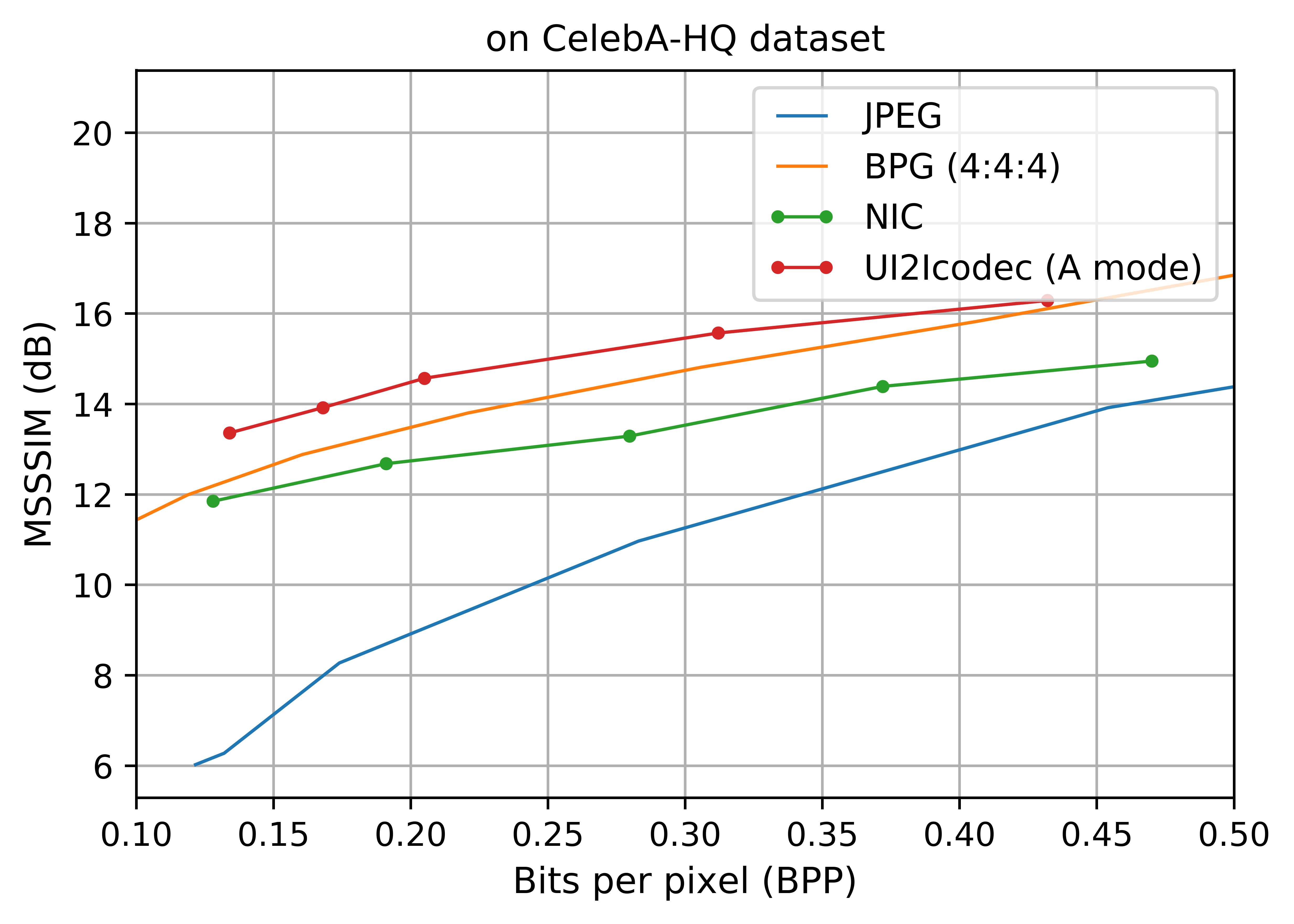}}
  \subfigure[LPIPS $\downarrow$]{\label{RDLPIPS_celebahq}\includegraphics[width=57mm]{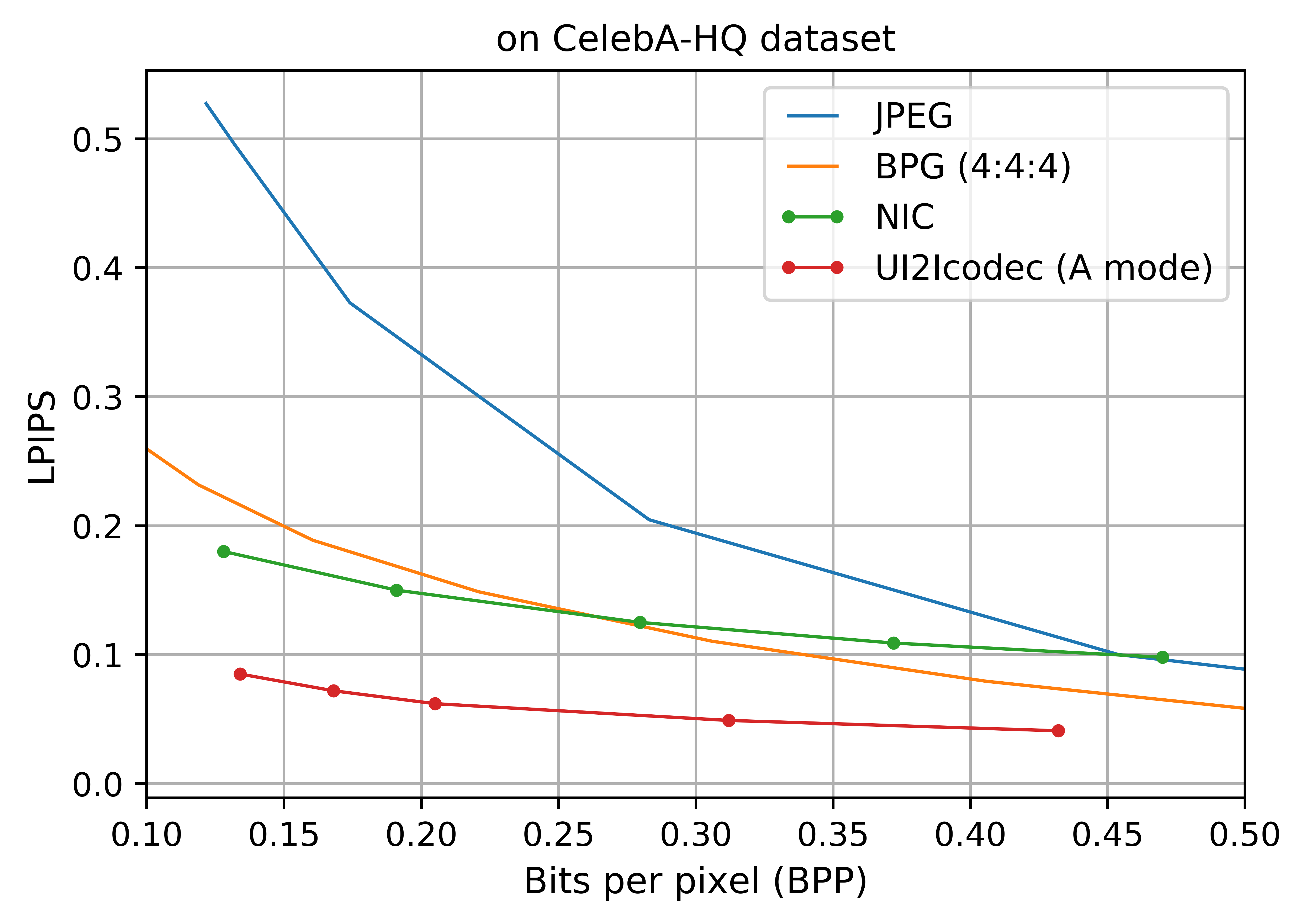}}  
  \caption{Rate-distortion and -perception curves on AFHQ and CelebA-HQ datasets.}
  \label{RDcomparison_afhp_celebahq}
\end{figure*}

\begin{figure*}[h!]
  \centering
  \includegraphics[width=0.9\textwidth]{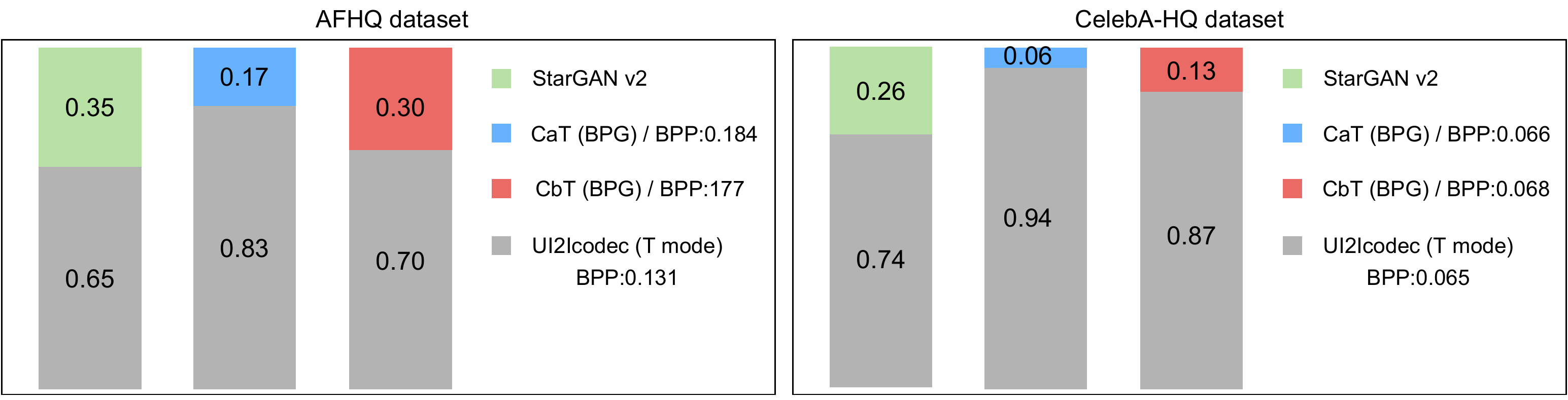}
  \caption{User study for the comparisons between StarGAN v2, CaT (BPG), CbT (BPG) and UI2Icodec (T mode) respectively on AFHQ and CelebA-HQ datasets.}
  \label{user_study}
\end{figure*}

\begin{figure*}[h!]
  \centering
  \subfigure[AFHQ dataset.]{\label{I2Icodec_a2}\includegraphics[width=85mm]{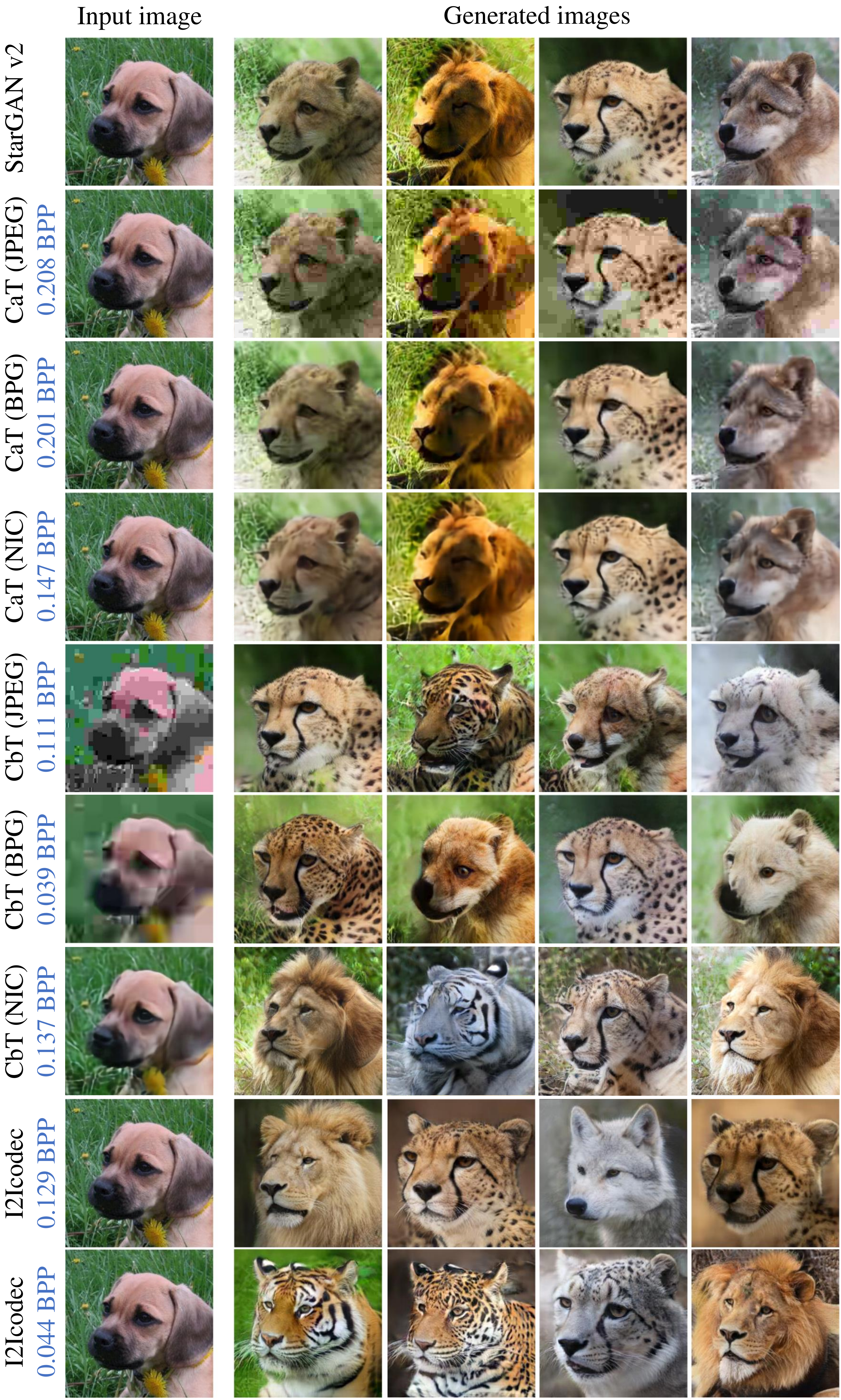}}
  \subfigure[CelebA-HQ dataset]{\label{I2Icodec_b2}\includegraphics[width=85mm]{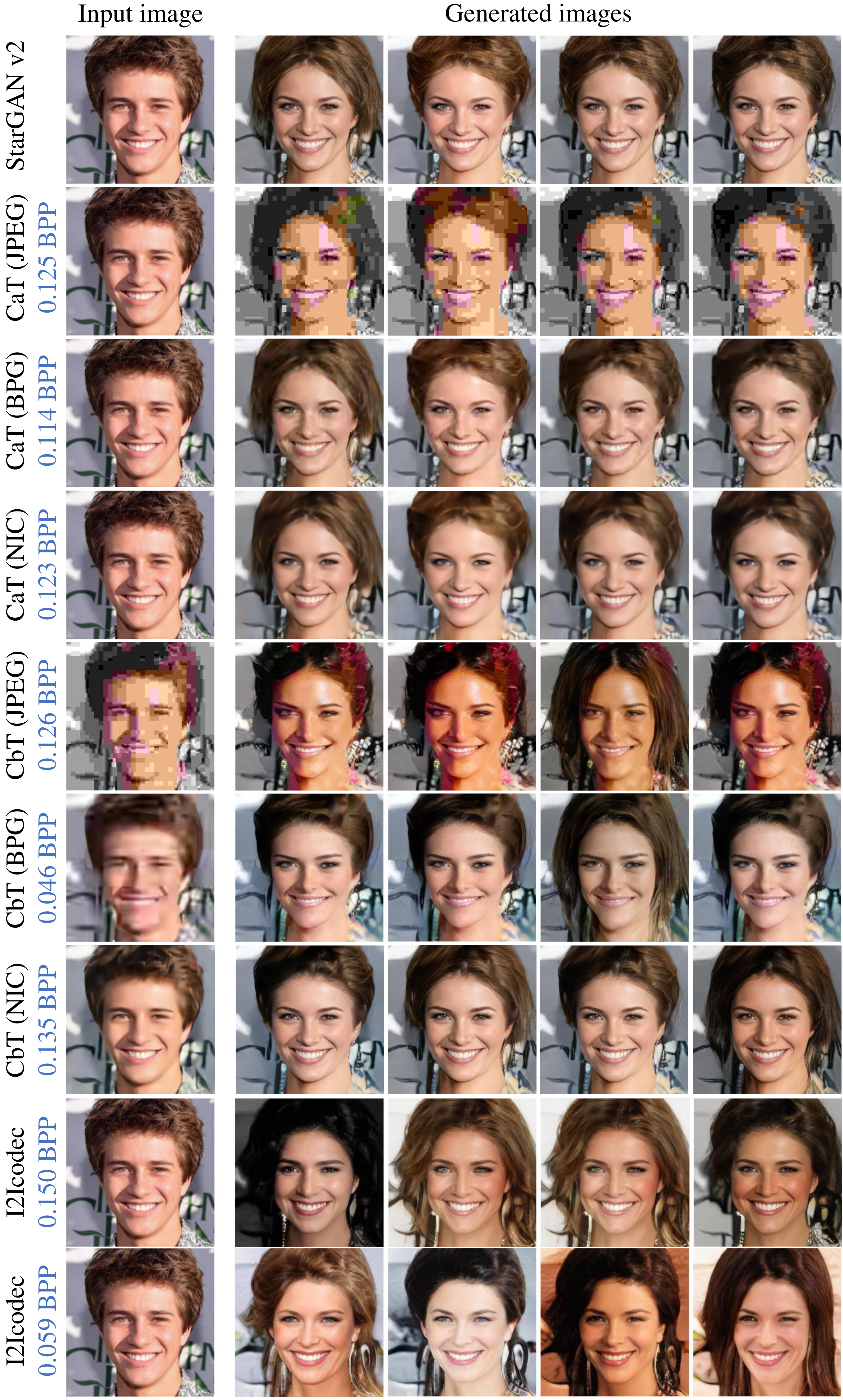}}
  \caption{Additional visualization of latent-guided synthesis with different methods.}
  \label{I2Icodec_vis2}
\end{figure*}

\subsection{Unified I2I translation and autoencoding}
\label{sec:UI2Icodec}
In this section, we evaluate the performance of UI2Icodec in I2I translation and autoencoding.

\textbf{Autoencoding.} We compare JPEG, BPG, and NIC as image compression baselines. As shown in Figure~\ref{RDcomparison_afhp_celebahq}, UI2Icodec in the autoencoding mode outperforms JPEG largely and BPG marginally on PSNR and MS-SSIM (dB) on similar rates. NIC has the best PSNR results since it was optimized for mean square error. In contrast, UI2Icodec obtains much better LPIPS scores than other methods due to the adversarial loss. Comparing the examples in Figure~\ref{rec_visualization}, we can see that our method can keep more high-frequency information and more natural reconstruction at the similar rate due to the leverage of GAN loss for autoencoding.

\textbf{I2I translation.} The quantitative evaluation of UI2Icodec in the translation mode is shown in Table.~\ref{I2I_performance}. It shows that it is possible to switch image compression and I2I translation by using our method. In addition, some images generated with UI2Icodec have been already shown in Figure~\ref{UI2Icodec_T_visualization}. More visualization samples can be viewed in Figure~\ref{UI2Icodec_h} and Figure~\ref{UI2Icodec_a}.

\subsection{Additional results}
\label{sec:ablation}
\textbf{Ablation study.} We evaluate the effects of the two main modifications of StarGAN v2 architecture: quantization+entropy coding (for compression), and adaptation units (for integrated translation+autoencoding). Comparing StarGAN v2 and I2Icodec in Table.~\ref{I2I_performance}, we observe that compression tends to improve FID and LPIPS. In contrast, comparing StarGAN and T+A (StarGAN with adaptation units and autoencoding loss), we observe that the combination of both functionalities has a small penalty in those metrics. However, an important caveat is that FID and LPIPS could be somewhat limited as evaluation metrics in this setting, and further research is required.

\textbf{User study.} In addition to quantitative comparison of I2I translation, we conducted a user study where we asked subjects to select which results they consider more realistic, given the target label and having the same pose as the input image. We apply pairwise comparisons (forced-choice) with 12 users (100 image pairs/user) for I2I translation. Experiments are performed on the AFHQ and Celeba-HQ datasets separately. Figure~\ref{user_study} shows that UI2Icodec (T mode), which runs at the silimar BPPs to CaT (BPG) and CbT (BPG), can obtain better scores than the main baseline methods StarGAN v2, CaT (BPG) and CbT (BPG) respectively.

\textbf{Model size and training time.} Table.~\ref{model_size_time} shows that the proposed models (both I2Icodec and UI2Icodec) only have 1\% more parameters than StarGANv2. Note that CbT and CaT with NICs require around twice the amount of parameters (since there are two encoders and two decoders). Training requires 8.6\%/48\% more time (for I2Icodec/UI2Icodec, respectively). Similarly, note that training CbT and CaT with NICs requires training a NIC model and StarGAN v2, so I2Icodec requires less training time. In addition, we need two independent models when both translation and autoencoding functionalities are needed (i.e. NIC and StarGAN), requiring 89.03M parameters, 65\% more than UI2Icodec (54.23M parameters). Finally, the overhead of UI2Icodec with respect to  I2Icodec is negligible (only around 0.01\%).

\begin{figure*}[htb]
  \centering
  \includegraphics[width=0.98\textwidth]{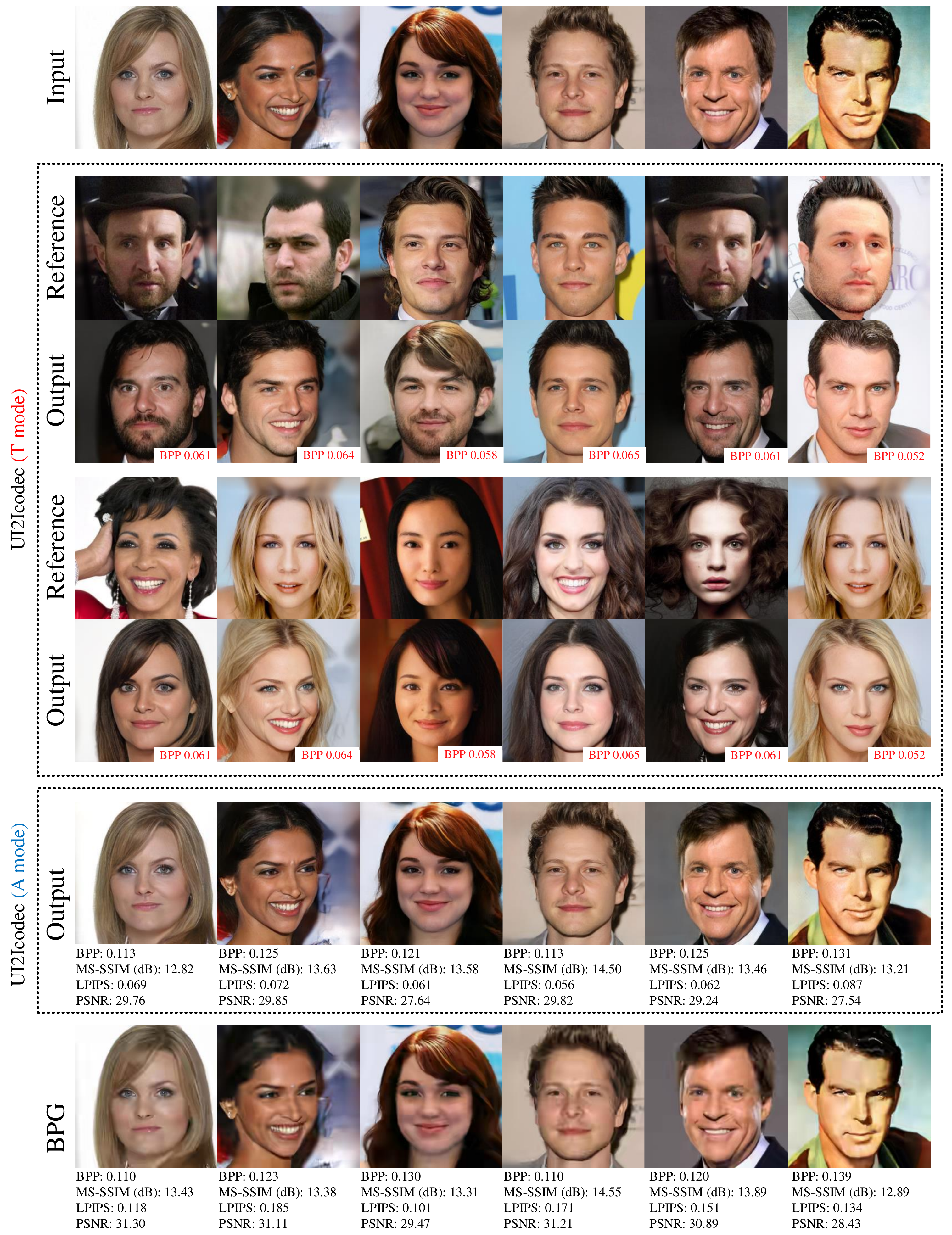}
  \caption{Diverse image synthesis results of UI2Icodec (reference-guided) in the translation and autoencoding modes on CelebA-HQ dataset.}
  \label{UI2Icodec_h}
  \vspace{-0.4cm}
\end{figure*}

\begin{figure*}[t]
  \centering
  \includegraphics[width=0.98\textwidth]{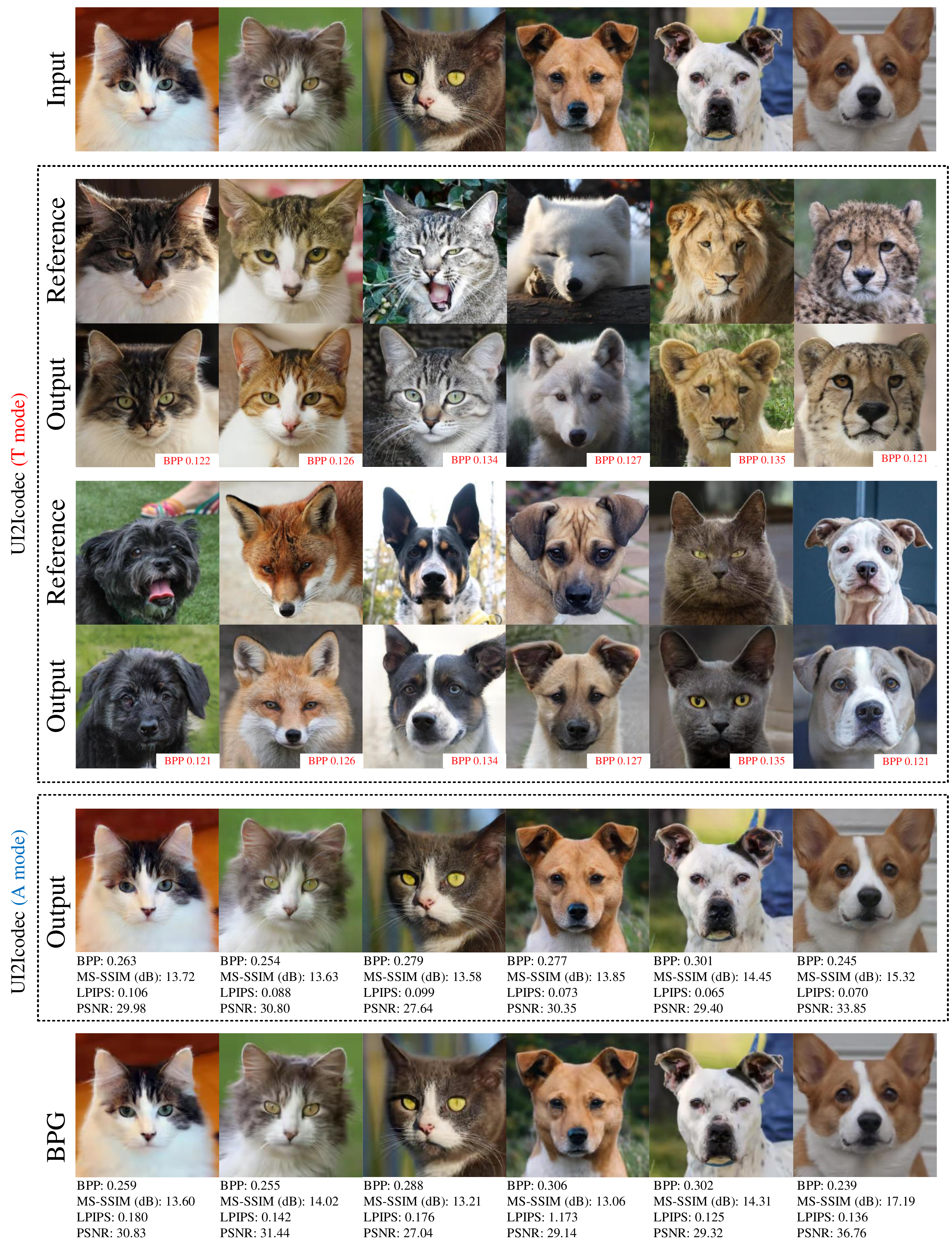}
  \caption{Diverse image synthesis results of UI2Icodec (reference-guided) in the translation and autoencoding modes on AFHQ dataset.}
  \label{UI2Icodec_a}
  \vspace{-0.4cm}
\end{figure*}

\textbf{More visualization results of I2Icodec and UI2Icodec.} We provide additional latent-guided image synthesis results from two independent I2Icodec models with different rate constraints (see eighth and ninth rows in Figure~\ref{I2Icodec_vis2}). Same with Figure~\ref{I2Icodec_vis}, we also include results of seven baselines: StarGAN v2, Compression after translation (CaT) with JPEG, BPG and NIC, Compression before translation (CbT) with JPEG, BPG and NIC. It shows that CaT and CbT suffer artifacts or result in unnatural or blurred translation, and our I2Icodec can generate natural and diverse images even with an extremely low rate. We also show more results of our UI2Icodec in both translation (\textcolor{red}{T}) and autoencoding (\textcolor{blue}{A}) modes on CelebA-HQ dataset (Figure~\ref{UI2Icodec_h}) and AFHQ dataset (Figure~\ref{UI2Icodec_a}) separately. It verifies that the UI2Icodec can successfully switch between modes using a single model. Our method can obtain better reconstructions than BPG, both visually and measured in terms of lower LPIPS values.

\section{Conclusion}
In this paper, we study the novel problem of distributed I2I translation, and its integration with autoencoding in a joint model, resulting in the proposed I2Icodec and UI2Icodec frameworks. Distributed I2I translation required augmenting an I2I translation framework with quantization and entropy coding. 
Interestingly, constraining the rate can control the amount of source information in I2I translation.
The experiments show that our joint model can keep competitive autoencoding and translation performance.

\section{Acknowledgments}
We acknowledge the support from Huawei Kirin Solution and the Spanish Government funding for projects RTI2018-102285-A-I00 and RYC2019-027020-I.

\bibliographystyle{elsarticle-num}
\bibliography{refs}

\end{document}